\documentclass[onecollarge,fleqn,runningheads]{svjour2}
\smartqed

\usepackage{graphicx}%
\usepackage{multirow}%
\usepackage{amsmath,amssymb,amsfonts}%
\usepackage{mathrsfs}%
\usepackage[title]{appendix}%
\usepackage{xcolor}%
\usepackage{textcomp}%
\usepackage{manyfoot}%
\usepackage{booktabs}%
\usepackage{algorithm}%
\usepackage{algorithmicx}%
\usepackage{algpseudocode}%
\usepackage{listings}%
\usepackage{upgreek}

\journalname{Theoretical and Computational Fluid Dynamics}

\begin{document}

\title{Eduction of acoustics-related coherent structures from Schlieren images of supersonic twin jets by coupling Doak's decomposition and SPOD}

\subtitle{}

\titlerunning{Acoustics-related coherent structures from Schlieren images of supersonic twin jets}

\author{Iv\'an Padilla-Montero \and Daniel Rodr{\'i}guez  \and Vincent Jaunet \and Peter Jordan
}

\institute{Iv\'an Padilla-Montero \and Daniel Rodr\'iguez \at ETSIAE-UPM, School of Aeronautics, Universidad Polit\'ecnica de Madrid, Pza. Cardenal Cisneros 3, 28040 Madrid, Spain \\
         \email{ivan.padilla@upm.es}           
\and Vincent Jaunet \and Peter Jordan \at D\'{e}partement Fluides, Thermique et Combustion, Institut Pprime, CNRS - Universit\'{e} de Poitiers - ISAE-ENSMA, 86036 Poitiers, France
}

\date{Received: date / Accepted: date}

\maketitle

\begin{abstract}
This work proposes a methodology to improve the extraction of coherent structures associated with the generation of acoustic fluctuations in turbulent jets from high-speed Schlieren images. This methodology employs the momentum potential theory of Doak~\cite{Doak1989,Jordan2013} to compute potential (acoustic and thermal) energy fluctuations from the Schlieren images by solving a Poisson equation, in the manner introduced by Prasad \& Gaitonde~\cite{Prasad2022}. The calculation of momentum potential fluctuations is then combined with the spectral proper orthogonal decomposition (SPOD) technique: the cross-spectral density is defined based on the momentum potential field, instead of the Schlieren images. While the latter are dominated by a broad range of vortical fluctuations in the turbulent mixing region of unheated high-speed jets, the momentum potential field is governed by acoustic fluctuations and its spatial structure in the frequency domain is remarkably coherent. This approach is applied here to Schlieren visualizations of a twin-jet configuration with a small jet separation and two supersonic operation conditions: a perfectly-expanded and a overexpanded one. 
The SPOD modes based on momentum potential fluctuations retain the wavepacket structure including the direct Mach-wave radiation together with upstream- and downstream-traveling acoustic waves, similar to SPOD modes based on the Schlieren images. However, they result in a remarkably lower-rank decomposition than Schlieren-based SPOD and, as opposed to the latter, provide an effective separation of twin-jet fluctuations into independent toroidal and flapping oscillations that are recovered as different SPOD modes.

\end{abstract}

\section{Introduction}\label{sec:intro}

Noise pollution remains one of the major environmental health concerns associated with aviation, owing to its significant impact on the activity of human and animal life. Jet-engine noise is one of the main contributors to this problem, affecting both urban environments with civil aviation operations as well as military environments such as launch centers or aircraft carriers. Supersonic multi-jet engines are frequently encountered in the propulsion systems of rocket launchers and modern high-speed aircraft. Among these, twin-jet configurations are the most common. Closely-spaced jets are known to strongly interact at the hydrodynamic and acoustic levels, giving rise to more complex flow structures than single round jets and distinct noise radiation patterns~\cite{Bhat:AIAA77,Kantola:JSV81}.

The link between the far-field sound radiated by high-speed jets and the turbulent fluctuations found in their core and mixing regions was soon recognized as a crucial point for the understanding of sound-generation mechanisms, and has been the subject of considerable research~\cite{JordanColonius:ARFM13}. Multiple studies on single round jets have shown that the radiated sound is highly directional for both subsonic and perfectly-expanded supersonic jets~\cite{CrightonHuerreJFM90,Tam:ARFM95,Cavalieri:JFM12}, and that it is associated with large-scale, low-frequency fluctuations found in the mixing region~\cite{Juve:JSV80,Cavalieri:JSV11b}. Such large-scale fluctuations, today known as coherent structures, were first observed by Crow \& Champagne~\cite{CrowChampagne:JFM71}. These structures resemble the instability waves that develop in harmonically-forced jets, prompting their modeling by means of linear stability theory~\cite{CrightonGaster:JFM76,MichalkePrAS1984} and introducing the term ``wavepackets'' to refer to them.

During the last two decades, a successful modeling of wavepackets in isolated round jets was achieved by means of linear stability calculations, and their relevance in the coherent dynamics was further demonstrated through experimental comparisons and high-fidelity simulations~\cite{Suzuki:JFM06,Gudmundsson:JFM11,Cavalieri:JFM13}. Vortex sheet/finite thickness models~\cite{Tam1989}, parabolized stability equations (PSE)~\cite{Piot:IJAAF06,Ray:PF09,Rodriguez2013,Sinha:JFM14} and, more recently, one-way Navier-Stokes equations~\cite{Towne:JCP15} and resolvent analysis~\cite{Garnaud:JFM2013,Jeun:PF2016,Schmidt:JFM2018} have been established as valid wavepacket modeling strategies. These studies have led to big advancements in the understanding of sound generation mechanisms through the modeling of Kelvin-Helmholtz instabilities and of acoustic resonances involving duct modes and shock-cell interactions in supersonic jets~\cite{Towne2017,Edgington-Mitchell2022}. 

Analogous wavepacket models for subsonic and supersonic twin jets have only been developed in the last few years, as they pose additional challenges in terms of complexity and computational cost compared to single round jets: the azimuthal Fourier decomposition of the flow field is no longer possible due to the loss of axial symmetry of the flow structure of each jet, thus requiring the use of three-dimensional techniques. Studies based on local cross-plane linear stability theory with two inhomogeneous directions~\cite{Rodriguez2018,Nogueira2021,Rodriguez2022arXiv}, along with plane-marching PSE~\cite{Rodriguez2021,PadillaMontero2023} and twin-jet vortex sheet/finite thickness models~\cite{Stavropoulos2023} have been used to characterize the instabilities that govern the twin-jet system and model the corresponding wavepackets.

Despite the achieved progress, current wavepacket models are not yet able to provide a complete description of the mechanics of sound generation in twin-jet systems. Linear models present limitations when mechanisms involving multiple interacting waves or coupling between the fluctuations of each jet are present. The interpretation of results is further complicated in configurations with small jet separations as the wavepackets become heavily deformed, departing from axial symmetry and complicating the identification of the oscillation modes. Therefore, hand-in-hand experimental investigations are necessary to validate the models as well as to characterize and provide physical understanding on mechanisms that have not yet been described computationally.

With the increasing ability to record and process high-resolution, high-speed experimental measurements, data-driven techniques introduced some decades ago have recently evolved and gained attention due to their potential to extract information from experimental data. In particular, the spectral proper orthogonal decomposition (SPOD)~\cite{Berkooz:ARFM93,Citriniti:JFM00,Towne2018} offers a powerful framework to extract coherent structures from time-resolved experimental visualizations, such as those obtained from high-speed Schlieren imaging. Recent investigations have employed this technique to study the screech resonance mechanism in single round jets with very satisfactory results~\cite{Edgington-Mitchell2022,Karnam2023}. For twin-jet configurations, the application of SPOD to experimental Schlieren data sets has also been successful to describe screech resonances~\cite{Nogueira2021b,Wong2023,PadillaMontero2023b}, where the associated coherent structures are composed of waves which, within each jet, feature a spatial structure analogous to a single azimuthal mode in the equivalent isolated jet. Conversely, at other frequencies where mixing-layer noise is the dominant mechanism, SPOD based on Schlieren images is not well suited for separating the toroidal and flapping modes of oscillation of the twin-jet system. As will be shown in this study, both types of oscillation often appear superposed in the same SPOD mode.

In a recent work, Prasad \& Gaitonde~\cite{Prasad2022} presented an approach to extract coherent structures from high-speed Schlieren images using a quantity derived from the momentum potential theory formulation of Doak~\cite{Doak1989}. Jordan et al.~\cite{Jordan2013} were the first to apply the momentum potential formulation to a flow problem, consisting of a solenoidal wavepacket, to investigate the source and flux terms involved in Doak's fluctuation-energy balance and their role in the generation of sound radiation. Later on, Unnikrishnan et al.~\cite{Unnikrishnan2016,Unnikrishnan2018} applied the technique to a supersonic cold jet flow field obtained by means of large-eddy simulation. In the approach of~\cite{Prasad2022}, the Helmholtz decomposition of the momentum density is used to derive a Poisson equation relating the Schlieren fluctuation field with the streamwise gradient of the momentum potential fluctuation field integrated along the line of sight. By applying SPOD to the derived momentum potential fluctuations, coherent structures representing the acoustic and thermal energy components of the flow may be obtained. In~\cite{Prasad2022}, this technique is applied to Schlieren visualizations of single round jets and twin rectangular jets, educing coherent structures associated with sound generation and with the waves responsible for resonant feedback loops. Nevertheless, the performance and limitations of this approach compared to the direct use of Schlieren images to extract coherent information have not yet been characterized. In particular, the impact of imposing \textit{ad hoc} boundary conditions for the solution of the Poisson equation in a truncated domain where the flow field at the boundaries is not purely irrotational remains unclear~\cite{Schoder2019,Schoder2020}.

In this work, the idea introduced by Prasad \& Gaitonde~\cite{Prasad2022} is adopted to improve the extraction of coherent structures from high-speed Schlieren measurements in twin supersonic jets. A new methodology is proposed to integrate the calculation of the streamwise gradient of the momentum potential field within the SPOD algorithm, paying special attention to the influence of the \textit{ad hoc} boundary conditions imposed on the potential field. The obtained coherent structures are compared against the SPOD structures provided by the direct use of Schlieren images, allowing a quantification of the benefits attained from the use of the derived momentum potential fluctuations for this problem.

The remainder of the paper is organized as follows. Section \ref{sec:exp_setup} presents the experimental twin-jet setup employed to obtain Schlieren visualizations. Section \ref{sec:methodology} describes the methodology used for the calculation of coherent structures, outlining the derivation of the Poisson equation relating Schlieren and momentum potential fluctuations, the SPOD algorithm and the integration of the calculation of the momentum potential field within it. Section \ref{sec:results} presents SPOD results for two different twin-jet operating conditions and highlights the differences between the coherent information obtained from momentum potential fluctuations against Schlieren fluctuations. Finally, concluding remarks are provided in \S~\ref{sec:conclusions}.

\section{Experimental setup} \label{sec:exp_setup}

\begin{figure}[t]
\centering
\includegraphics[width=0.85\textwidth]{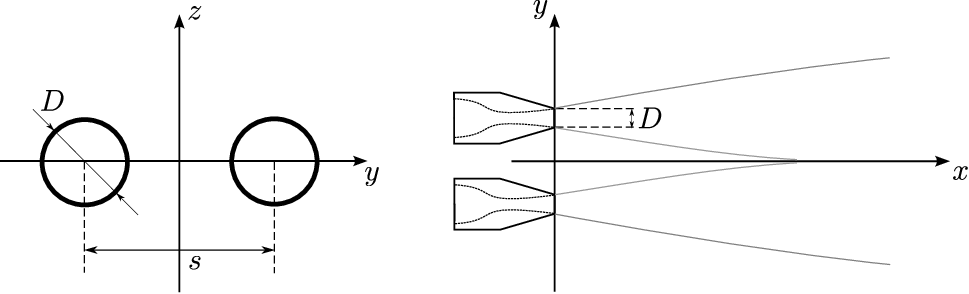}
\caption{Twin jet geometry and associated parameters.}
\label{fig:geometry}
\end{figure}

\begin{figure}[t]
\centering
\includegraphics[width=0.99\textwidth]{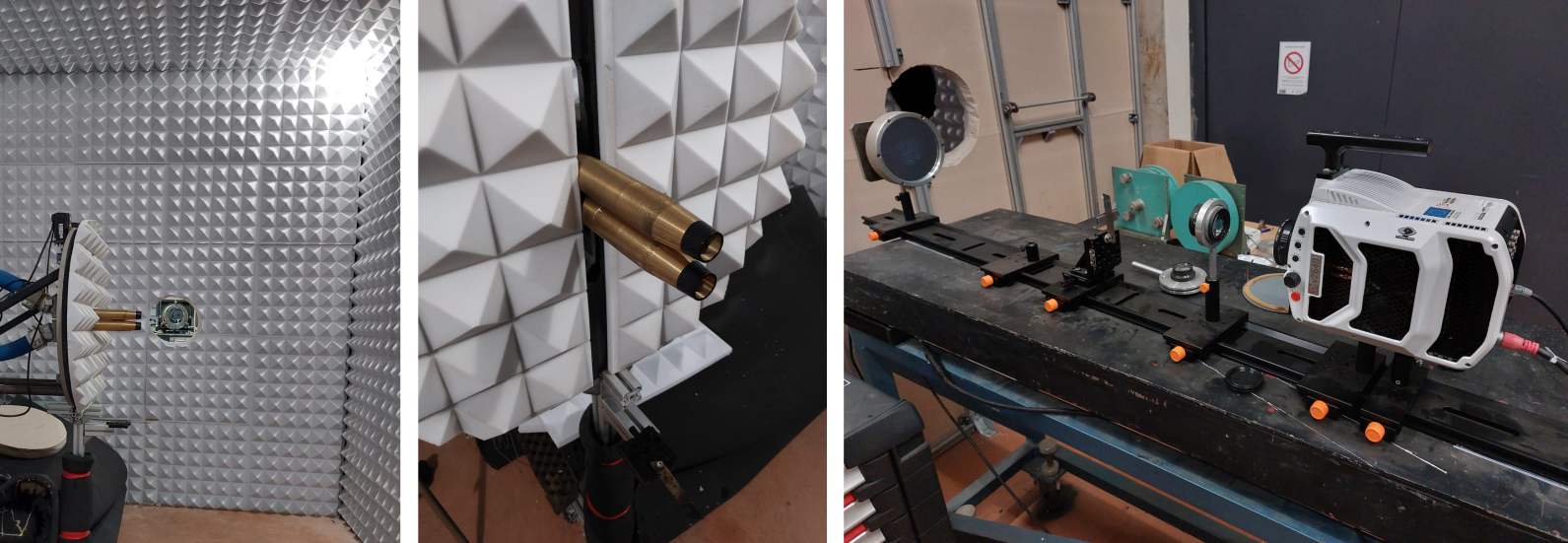}
\caption{Experimental setup: (left) overall view of the twin-jet system; (middle) close-up view of the twin nozzles; (right) some elements of the Schlieren system: flat mirror, knife edge, lens and camera.}
\label{fig:experimental_setup}
\end{figure}

The studied twin-jet configuration is represented in figure \ref{fig:geometry}. The jets are generated by two identical round convergent-divergent nozzles. The nozzle geometry has been designed at Institut Pprime (CNRS-Universit\'{e} de Poitiers-ISAE-ENSMA), and follows a truncated ideal contour (TIC) profile with an exit diameter of $D = 0.025$ m and an exit-to-throat area ratio of $A_e/A_t = 1.225$. The center of each of the nozzles is located along the $y$-axis ($z = 0$) and the exit of each of them is located at $x = 0$. The spacing between the nozzle's axes is denoted by $s$.

The experimental Schlieren visualizations have been performed at the PROM\'{E}T\'{E}E platform of Institut Pprime (CNRS-Universit\'{e} de Poitiers-ISAE-ENSMA). The facility employed is the T200 compressible wind tunnel, which is powered by a $200$ bar compressed air network and can reach operational conditions up to an isentropic Mach number $M_j = 2$ for the employed nozzle geometry. A heating system based on a series of tanks with heated nickel balls is used to increase and maintain the total temperature of the air arriving to the nozzles. The first and second pictures shown in figure \ref{fig:experimental_setup} display the twin-jet experimental system built in the facility, which is placed inside a semi-anechoic room. The jet spacing considered for this study is $s/D = 1.76$, which is the minimum possible nozzle spacing according to the outer surface dimensions of the designed nozzles.

The jets are subject to different nozzle pressure ratios (NPR) with a fixed total temperature of $T_{t0} = 300$ K.  
The pressure ratio considered, i.e. the ratio of the total pressure in the reservoir $p_{t0}$ to the ambient pressure $p_\infty$ is defined in terms of the isentropic jet Mach number $M_j$ through the isentropic relation $p_{t0} / p_\infty = (1 + 0.5(\gamma - 1) M_j^2)^{(\gamma - 1)/\gamma}$. 
Since the jets are not heated, the flow acceleration within the nozzle results in jet static temperatures lower than the ambient temperature. The operation condition closest to the perfectly-expanded regime that could be achieved in practice is obtained at $M_j = 1.54$, calibrated experimentally by means of flow visualization. In this work, two different operating conditions are considered, namely $M_j = 1.54$ and $M_j = 1.26$, the latter corresponding to an overexpanded regime. The corresponding Reynolds number based on the nozzle exit diameter and the jet exit flow conditions for $M_j = 1.54$ is $Re =  1.40 \times 10^6$.

\begin{figure}[t]
\centering
\includegraphics[width=0.99\textwidth]{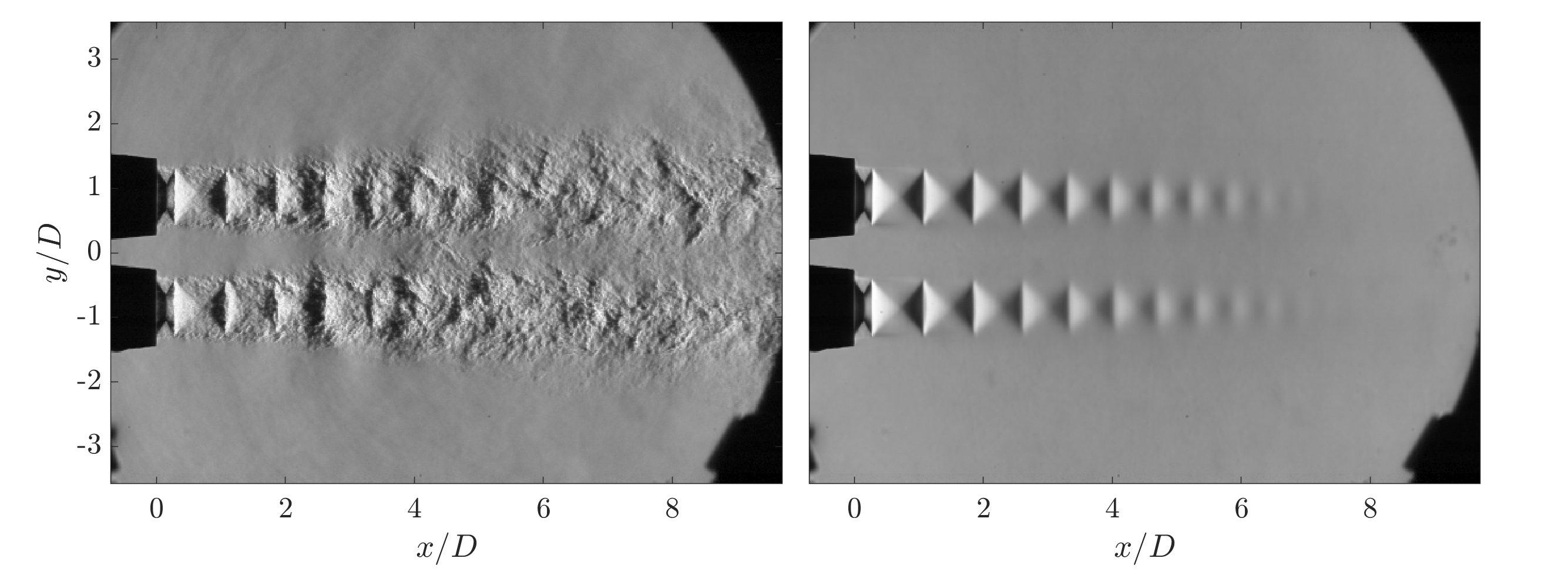}
\caption{Schlieren visualizations of the twin jet at $M_j = 1.26$ (overexpanded regime): (left) sample instantaneous snapshot; (right) mean field obtained by averaging 30,000 snapshots.}
\label{fig:Schlieren_Mj1d26}
\end{figure}

The optical system consists of a classical Z-type Schlieren setup. Some of its elements are shown in the third picture of figure \ref{fig:experimental_setup}. A continuous light source is provided by a 60 W LED, which goes through an aperture that prevents any light from the source to directly enter the test section. Two parabolic mirrors with a diameter of 30 cm and a focal length of 3 m are used to produce a collimated light beam which, through two circular apertures, traverses the test section along the $z$ direction according to the reference frame shown in figure \ref{fig:geometry}. Two additional flat mirrors with a diameter of 12 cm each are used to increase the path length of the system while keeping it in a reduced space. A vertical knife edge is placed at the focal length of the second parabolic mirror; in consequence, the light intensity field in the recovered Schlieren images is proportional to the streamwise density gradients in the flow ($\partial \rho / \partial x$).

The camera employed to record the images at high-speed is a Phantom v2640. For each operating condition, 30,000 snapshots are recorded at a sampling frequency of $f_s = 68$ kHz with a spatial resolution of $352 \times 512$ pixels. An exposure time 0.8 $\upmu$s is used, which is small enough to freeze the convected flow disturbances while ensuring enough contrast in the image. The recorded images consist of an array of pixel intensity values ranging between 0 and 4095 (12-bit color depth).

The image's spatial window approximately comprises an area of $3D$ away from each nozzle axis along the $y$ direction and $10D$ downstream of the nozzle exit along the $x$ direction. This size was found to offer a good compromise between spatial resolution and window size for the current setup, resulting in a pixel size of $0.02D$. Figure \ref{fig:Schlieren_Mj1d26} displays an instantaneous Schlieren snapshot obtained at $M_j = 1.26$, as well as the corresponding mean field obtained by averaging all 30,000 snapshots.

\section{Methodology} \label{sec:methodology}

In the following, instantaneous flow quantities $q$ are split into a time-stationary mean component and a time-dependent fluctuating component: $q(x,y,z,t) = \bar{q}(x,y,z) + q'(x,y,z,t)$.

\subsection{Doak's momentum potential theory}
\label{sec:Doak}

The momentum potential formulation proposed by Doak \cite{Doak1989} is based on a Helmholtz decomposition of the form:

\begin{equation}
\rho u_i = B_i - \frac{\partial \psi}{\partial x_i},
\label{eq:Doak}
\end{equation}

\noindent where $\rho u_i$ denotes the momentum density of the flow (density $\rho$ times velocity $u_i$), $B_i$ is the solenoidal momentum component ($\partial B_i / \partial x_i = 0$) and $\psi$ denotes the scalar potential associated with the irrotational momentum component ($\nabla \times \partial \psi / \partial x_i = 0$).

Introducing the decomposition \eqref{eq:Doak} into the continuity equation

\begin{equation}
\frac{\partial \rho}{\partial t} + \frac{\partial \left( \rho u_i \right)}{\partial x_i} = 0,
\end{equation}

\noindent assuming a statistically time-stationary flow ($\partial \bar{\rho} / \partial t = 0$), and noting that $\rho(x_i,t) = \bar{\rho}(x_i) + \rho'(x_i,t)$ and $\psi(x_i,t) = \bar{\psi}(x_i) + \psi'(x_i,t)$, yields, on one hand:

\begin{equation}
\frac{\partial^2 \psi'}{\partial x_i^2} = \frac{\partial \rho'}{\partial t},
\label{eq:Poisson}
\end{equation}

\noindent which is a Poisson equation relating the two scalar fluctuation fields $\rho'$ and $\psi'$, and, on the other hand:

\begin{equation}
\frac{\partial^2 \bar{\psi}}{\partial x_i^2} = 0,
\end{equation}

\noindent which implies that the scalar potential field has zero mean, i.e. $\psi(x_i,t) = \psi'(x_i,t)$ (see \cite{Doak1989}).

Helmholtz's decomposition theorem expressed in the form of equation \eqref{eq:Doak} holds for vector fields that decay to zero sufficiently fast at infinity~\cite{Batchelor2000,Sprossig2010}. This requirement is usually not satisfied in the finite domains employed for the numerical simulation or experimental visualization of flow fields of practical interest, such as the twin-jet configuration under study. Certainly, the momentum density of the twin-jet flow field does not vanish at the boundaries of the domain shown in figure~\ref{fig:Schlieren_Mj1d26}. In these cases, the truncation of the domain introduces a third component in the decomposition which corresponds to an harmonic solution component (see Theorem 2 in~\cite{Schoder2020}):

\begin{equation}
\rho u_i = B_i - \frac{\partial \psi}{\partial x_i} - H_i.
\end{equation}

\noindent The harmonic component, denoted by $-H_i$, is both solenoidal and irrotational~\cite{Batchelor2000}. As a result, it also has an associated scalar potential $H_i = \partial \psi_h / \partial x_i$ that satisfies Laplace's equation ($\nabla \cdot H_i = \partial^2 \psi_h / \partial x_i^2 = 0$) in the domain of analysis. This component represents the impact of the truncation of the domain on the Helmholtz decomposition~\cite{Schoder2019,Schoder2020}. Consequently, as the Schlieren images are contained in a truncated domain with boundaries at which the flow momentum does not vanish, $H_i$ will also be recovered as part of the solution of equation \eqref{eq:Poisson}. In those cases where \textit{ad hoc} boundary conditions are enforced in the solution of the Poisson equation, it is crucial to pay attention to the impact that the harmonic field has on the physically-relevant part of the solution, as discussed in section \ref{sec:Poisson_solution}.

\subsection{Relation between Schlieren fluctuations and scalar momentum potential fluctuations}

According to the experimental set-up described in section \ref{sec:exp_setup}, the recorded Schlieren images represent streamwise density gradients integrated along the line of sight ($z$ axis), that is

\begin{equation}
\sigma (x,y) = \int \frac{\partial \rho (x,y,z)}{\partial x}\,dz.
\end{equation}

Following Prasad \& Gaitonde \cite{Prasad2022}, equation \eqref{eq:Poisson}, owing to its linear nature, can be differentiated with respect to $x$ to obtain:

\begin{equation}
\frac{\partial^2}{\partial x_i^2} \left( \frac{\partial \psi'}{\partial x} \right) = \frac{\partial }{\partial t} \left( \frac{\partial \rho '}{\partial x} \right).
\label{eq:Poisson_dx}
\end{equation}

\noindent If integrated along the Schlieren line of sight, equation \eqref{eq:Poisson_dx} directly relates the fluctuation of the Schlieren field with the streamwise derivative of the potential fluctuation field $\psi'$ integrated along the line of sight, that is

\begin{equation}
\frac{\partial^2 \Theta'}{\partial x_i^2} = \frac{\partial \sigma'}{\partial t},
\label{eq:Poisson_sch}
\end{equation}

\noindent where $\sigma' = \int (\partial \rho' / \partial x)\,dz$ denotes the Schlieren fluctuation field and $\Theta' = \int (\partial \psi' / \partial x)\,dz$.

As shown by Doak \cite{Doak1989}, the fluctuating potential $\psi'$ can be written as the sum of acoustic and thermal components, $\psi' = \psi'_A + \psi'_T$, which are directly related with the pressure and the entropy fluctuation fields, respectively. This decomposition bears strong resemblance with the generalized potential disturbance energy definition introduced by Chu \cite{Chu1965}, which consists of an acoustic component associated with compression due to pressure fluctuations, and a thermal component associated with heat exchange due to entropy spottiness. In this regard, $\psi'$, and by extension $\Theta'$, can be interpreted as quantities describing the potential fluctuation energy in the system. Previous works \cite{Unnikrishnan2016,Unnikrishnan2018} have shown that the thermal component is non-radiating in cold supersonic turbulent jets. As a result, outside of the jet core, $\Theta'$ represents the acoustic energy present in such flows.

Equation \eqref{eq:Poisson_sch} may be interpreted as a means to calculate $\Theta'$ from experimental Schlieren measurements. However, for this statement to be rigorous, the boundary conditions imposed on equation \eqref{eq:Poisson_sch} must satisfy the irrotationality condition inherent in $\partial \psi'/ \partial x_i$. In general, this is not possible when dealing with experimental datasets, as the calculation of the boundary values of $\Theta'$ would require the integration of equation \eqref{eq:Doak} along far-field boundaries where the momentum density of the flow is purely irrotational and in consequence $\rho u_i = - \partial \psi' / \partial x_i$. This condition can only be rigorously satisfied in certain cases such as the solenoidal wavepacket investigated by Jordan et al.~\cite{Jordan2013}. Despite this difficulty, the solution of the Poisson equation can still be carried out in practice using an \textit{ad hoc} treatment at the domain boundaries, at the expense of introducing unphysical waves in the solution through the harmonic component (see section \ref{sec:Doak}). This is, to the best of the author's knowledge, the approach adopted by~\cite{Unnikrishnan2016,Unnikrishnan2018,Prasad2022}.

In this work, the proposed methodology enables the identification of the harmonic solution component that arises from the domain truncation and associated boundary conditions, and allows to remove its influence from the physically-relevant solution component, as described in section \ref{sec:Poisson_solution}.

\subsection{Spectral proper orthogonal decomposition}

The spectral proper orthogonal decomposition (SPOD) algorithm as described by Towne et al. \cite{Towne2018} is employed here to extract the spatio-temporal coherent structures from the time-resolved experimental data. The algorithm employs Welch's method to average the spectral information over multiple realizations of the flow and as a result obtain convergent estimates of the cross-spectral density tensor. In this study, the number of realizations in use is much smaller than the number of grid points (image pixels) of the Schlieren data sets, and therefore, for a given frequency, the snapshot method \cite{Sirovich:QAM87} variant of the SPOD eigenvalue problem is used:

\begin{equation}
\tilde{\mathbf{Q}}_k^H \mathbf{W} \tilde{\mathbf{Q}}_k \mathbf{\Phi}_k = \mathbf{\Phi}_k \mathbf{\Lambda}_k,
\label{eq:SPOD_EVP}
\end{equation}

\noindent where $\tilde{\mathbf{Q}}_{k}$ is the matrix whose columns contain the temporal discrete Fourier transform (DFT) of each segment of the total time-series used in Welch's averaging (considered as an independent realization) at the $k$th frequency, $\mathbf{W}$ is the matrix that contains the weight and the numerical quadrature defining the SPOD spatial inner product, $\mathbf{\Phi}_k$ is a matrix whose columns contain the eigenvectors of the problem for the $k$th frequency and $\mathbf{\Lambda}_k = \mbox{diag}(\lambda_1,\lambda_2,\dots)$ is a diagonal matrix containing the associated eigenvalues for that particular frequency. The superscript $H$ denotes the complex-conjugate transpose. The eigenvalues found in $\mathbf{\Lambda}_k$ directly represent the power spectral density associated with each of the SPOD modes. The SPOD modes for the $k$th frequency, denoted by $\mathbf{\Psi}_k$, are obtained by means of the following relation:

\begin{equation}
\mathbf{\Psi}_{k} = \tilde{\mathbf{Q}}_{k} \mathbf{\Phi}_{k} \mathbf{\Lambda}_{k}^{-1/2}.
\end{equation}

The assembly of the matrix $\tilde{\mathbf{Q}}_{k}$ for each frequency requires prior calculation of the temporal DFT of each realization of the flow. In this case, each realization consists of a data matrix (block) with a segment of the time-series of Schlieren fluctuation  ($\sigma'$) or $\Theta'$ snapshots of equal length. In this work, each data set consists of 30,000 instantaneous snapshots, which are divided into 57 blocks of 1024 snapshots each with a 50\% overlap. Each block is windowed using a Hamming window to reduce spectral leakage.

Given the nature of the experimental data sets (pixel intensities), the definition of an integral energy norm is not meaningful. Instead, a weighted 2-norm based on a trapezoidal integration rule (see \cite{Schmidt2020}) is employed to serve as a numerical quadrature for the SPOD inner product.

This work focuses on those coherent structures that describe the spatio-temporal dynamics of the twin jet system when both jets are coupled, that is, structures that involve both jets. For this purpose, the symmetry of the system is exploited to generate symmetric and antisymmetric fluctuation fields. When Schlieren realizations are considered, each snapshot is split by the line at $y = 0$ to create two new data sets, namely, a symmetric data set $\sigma_s = (\sigma_u + \sigma_l)/2$ and an antisymmetric one $\sigma_a = (\sigma_u - \sigma_l)/2$, with $\sigma_u$ and $\sigma_l$ respectively denoting the upper ($y > 0$) and lower ($y < 0$) halves of the original Schlieren images. The SPOD is then applied to both the symmetric and antisymmetric data sets separately.

In the results presented in this work, the SPOD problem \eqref{eq:SPOD_EVP} is solved both in terms of the cross-spectral density of $\sigma'$ or in terms of the cross-spectral density of $\Theta'$. This is done by constructing the $\tilde{\mathbf{Q}}_k$ matrix using either the temporal DFT of Schlieren realizations or the temporal DFT of $\Theta'$ realizations. The DFT of Schlieren realizations is straightforward to obtain by means of the fast Fourier transform (FFT). The approach employed to obtain the DFT of $\Theta'$ fields for each realization is described in the following section.

\subsection{Calculation of $\Theta'$ and its integration within the SPOD framework}
\label{sec:Poisson_solution}

Poisson equation \eqref{eq:Poisson_sch} can be solved directly in the space-time domain by employing discretizations of the temporal and spatial derivatives and solving a linear system of equations, as done by e.g. Prasad \& Gaitonde \cite{Prasad2022}. In this work, an alternative procedure is employed which solves the equation in the frequency-wavenumber domain by performing discrete Fourier transforms (DFT) in time and on the cartesian two-dimensional domain of the Schlieren images.

On one hand, solving for $\Theta'$ in the wavenumber domain prevents discretizing the spatial derivatives of the Laplacian operator and the consequent solution of a linear system of equations, therefore reducing significantly the computational cost of the methodology.
On the other hand, working in the frequency domain makes the calculation of the time derivative of the Schlieren fluctuations straightforward, avoiding the complications associated with the computation of time derivatives: as shown in Appendix \ref{sec:AppB}, computing the time derivative of $\sigma'$ by means of a low-order scheme results in a strong damping of the high-frequency content present in the Schlieren snapshots. Conversely, high-order differentiation schemes can present numerical instabilities and require of low-pass filters, resulting in added computation costs. 

In addition, the use of $\Theta'$ realizations in the SPOD algorithm requires the temporal Fourier transform of the $\Theta'$ field to construct the $\tilde{\mathbf{Q}}_k$ matrix. As a consequence, obtaining $\Theta'$ directly in the frequency domain does not involve additional cost, as it only brings forward the necessary temporal DFT.

Performing the temporal DFT, equation \eqref{eq:Poisson_sch} becomes:

\begin{equation}
\frac{\partial^2 \tilde{\Theta}'}{\partial x_i^2} = \mathrm{i} \omega \tilde{\sigma}',
\label{eq:Poisson_sf}
\end{equation}

\noindent where $\omega = 2 \pi f$ is the angular frequency and the tilde symbol denotes Fourier-transformed quantities in time, e.g. $\tilde{q}'(x,y,\omega)$. Performing the DFT in both $x$ and $y$, the resulting form of the equation in the frequency-wavenumber domain is

\begin{equation}
- \left( k_x^2 + k_y^2 \right) \hat{\Theta}' = \mathrm{i} \omega \hat{\sigma}',\quad \text{or} \quad \hat{\Theta}' = -\frac{\mathrm{i} \omega \hat{\sigma}'}{k_x^2 + k_y^2},
\label{eq:Poisson_kf}
\end{equation}

\noindent where $k_x$ and $k_y$ respectively denote the wavenumbers along $x$ and $y$ and the hat symbol refers to Fourier-transformed quantities in time and space, $\hat{q}'(k_x,k_y,\omega)$. Equation \eqref{eq:Poisson_kf} provides an algebraic expression to directly compute $\hat{\Theta}'$ for each frequency and wavenumber pair. In order to obtain $\tilde{\Theta}'$, which is the quantity of interest for the SPOD algorithm, the inverse DFT of $\hat{\Theta}'$ in space is carried out.

The aforementioned procedure imposes periodicity of the solutions at the boundaries of the Schlieren domain. In order to artificially place the periodic boundaries further away from the actual boundaries dictated by the Schlieren image size, the temporal Fourier-transformed Schlieren fluctuation $\tilde{\sigma}'$ is zero-padded along the $x$ and $y$ directions before transforming it to the wavenumber domain. To prevent a discontinuous jump in $\tilde{\sigma}'$, a Planck-taper spatial window \cite{McKechan2010} is used, which enables a smooth transition to zero at the domain boundaries without altering the interior values. The values of the window function $w$ for $N+1$ samples can be expressed as follows:

\begin{equation}
\begin{cases}
w_0 = 0,\\
w_j = \left[ 1 + \exp \left( \frac{\varepsilon N}{j} - \frac{\varepsilon N}{\varepsilon N - j} \right) \right]^{-1}, & \text{if } 1 \leq j < \varepsilon N\\
w_j = 1, & \text{if } \varepsilon N \leq j \leq N/2\\
w_j = w_{N-j}, & \text{if } N/2 < j \leq N
\end{cases}
\end{equation}

\noindent where $0 < \varepsilon \leq 0.5$ is the so-called tapering parameter. Then, after computing $\hat{\Theta}'$ in the wavenumber domain and inverting the spatial DFT, the spatial window is reduced back to the original one. 

As introduced before, the calculation of $\tilde{\Theta}'$ from Schlieren snapshots by means of the described methodology introduces harmonic unphysical solution components in the obtained $\tilde{\Theta}'$ fields due to the violation of the irrotationality condition at the boundaries of the Schlieren domain. The harmonic unphysical component is associated with the imposition of \textit{ad hoc} boundary conditions (in this case, periodic boundary conditions) in the artificially truncated physical domain; it is found to be sensitive to changes in the domain size and its energy to be concentrated in small streamwise wavenumbers, featuring streamwise phase speeds $c_{ph} = \omega/k_x$ that are highly supersonic and do not represent physically-sound components of the twin-jet system. Appendix \ref{sec:AppA} illustrates the signatures of these unphysical harmonic waves, and how they can be clearly distinguished from physically-meaningful components. No unphysical energetic components have been observed in the $y$ direction, which is attributed to the fact that for this problem, the main rotational flow regions at the boundaries are found at the streamwise boundaries (mainly the downstream boundary of the Schlieren images).

In order to prevent these harmonic components from contaminating the SPOD modes to be computed subsequently, the amplitude of those $\hat{\Theta}'$ components with a streamwise wavenumber $k_x$ that is smaller than a given cut-off value is set to zero. Such cut-off value is frequency dependent, and is fixed by specifying a supersonic streamwise phase speed $c_{ph,c}$ threshold. This filtering procedure is applied in the wavenumber domain, just before inverting the spatial DFT to obtain a $\tilde{\Theta}'$ fluctuation field that does not contain the energetic part of the unphysical harmonic component.

The proposed approach to construct the SPOD eigenvalue problem \eqref{eq:SPOD_EVP} based on $\Theta'$ can be summarized in the following steps:

\begin{enumerate}
\item Compute the temporal FFT of Schlieren fluctuation snapshots ($\tilde{\sigma}'$) for each realization (block).
\item For each frequency and each block:
\begin{enumerate}
	\item Add a zero-padding along $x$ and $y$ to the $\tilde{\sigma}'$ field, using a two-dimensional Planck-taper window. In this work, a tapering parameter $\varepsilon = 0.1$ has been employed. The SPOD solution was found to be very insensitive to $\varepsilon$ in the range $0 < \varepsilon < 0.15$.
	\item Compute the spatial FFT of the zero-padded $\tilde{\sigma}'$ field along $x$ and $y$.
	\item Calculate the $\hat{\Theta}'$ field using equation \eqref{eq:Poisson_kf}.
	\item Remove the most energetic unphysical harmonic waves by setting to zero the region of the $\hat{\Theta}'$ field comprised between $-\omega/c_{ph,c} < k_x < \omega/c_{ph,c}$, where $c_{ph,c}$ is the chosen phase-speed threshold. In this study, a value of $c_{ph,c} = 1.2 c_\infty$ has been used, where $c_\infty$ denotes the freestream speed of sound. The sensitivity of the SPOD results to the choice of $c_{ph,c}$ is reported in Appendix \ref{sec:AppA}.
	\item Compute the inverse spatial FFT of the filtered $\hat{\Theta}'$ field to obtain $\tilde{\Theta}'$.
	\item Reduce the spatial window of the $\tilde{\Theta}'$ field back to the original size of the Schlieren snapshots.
	\item If desired, create symmetric and antisymmetric $\tilde{\Theta}'$ fields as $\tilde{\Theta}'_s = (\tilde{\Theta}'_u + \tilde{\Theta}'_l)/2$ and $\tilde{\Theta}'_a = (\tilde{\Theta}'_u - \tilde{\Theta}'_l)/2$.
\end{enumerate}
\item For each frequency, build the $\tilde{\mathbf{Q}}_k$ matrix using the computed $\tilde{\Theta}'$ fields for all realizations.
\end{enumerate}

\subsection{Calculation of Schlieren SPOD modes for the $\Theta'$ cross-spectral density}
\label{sec:Sch_reconstruction}

The snapshot formulation of POD allows the computation of flow-field variables, corresponding to the SPOD modes, that are not involved in the definition of the cross-spectral density matrix~\cite{FreundColonius:AIAA02,Boree2003,Sinha:JFM14,Souza:AESCTE15,Souza:JFM19,Kaplan2021,Karban2022,Karban2023}. In particular, it allows the computation of the Schlieren fluctuation field $\sigma'$ (the line-of-sight integrated $\partial \rho' / \partial x$) associated with the SPOD modes corresponding to the cross-spectral density of $\Theta'$.

This is accomplished using the eigenvectors and eigenvalues of the $\Theta'$ decomposition ($\mathbf{\Phi}_{k,\Theta'}$ and  $\mathbf{\Lambda}_{k,\Theta'}$, respectively) combined with the matrix of frequency-domain realizations for the Schlieren fields ($\tilde{\mathbf{Q}}_{k,\sigma'}$):

\begin{equation}
\mathbf{\Psi}_{k,\sigma'} = \tilde{\mathbf{Q}}_{k,\sigma'} \mathbf{\Phi}_{k,\Theta'} \mathbf{\Lambda}_{k,\Theta'}^{-1/2}.
\label{eq:extended_SPOD}
\end{equation}

\noindent Note that this procedure is equivalent to solving an augmented SPOD eigenvalue problem consisting of a $\tilde{\mathbf{Q}}_k$ matrix containing both Schlieren and $\Theta'$ realizations, and then using a weight matrix that is non-zero only for the $\Theta'$ components.

\section{Results}\label{sec:results}

This section presents results obtained by applying the proposed methodology to the twin-jet Schlieren visualizations acquired for two different operating conditions: $M_j = 1.54$, corresponding approximately to a perfectly-expanded jet condition, and $M_j = 1.26$, corresponding to an over-expanded jet condition. SPOD results for the two definitions of the cross-spectral density matrix are presented. SPOD results based directly on the Schlieren images ($\sigma'$) are denoted as $CS$, while those based on $\Theta'$ are denoted by $C\Theta'$.

\subsection{Perfectly-expanded condition}

\begin{figure}[t]
\centering
\includegraphics[width=0.99\textwidth]{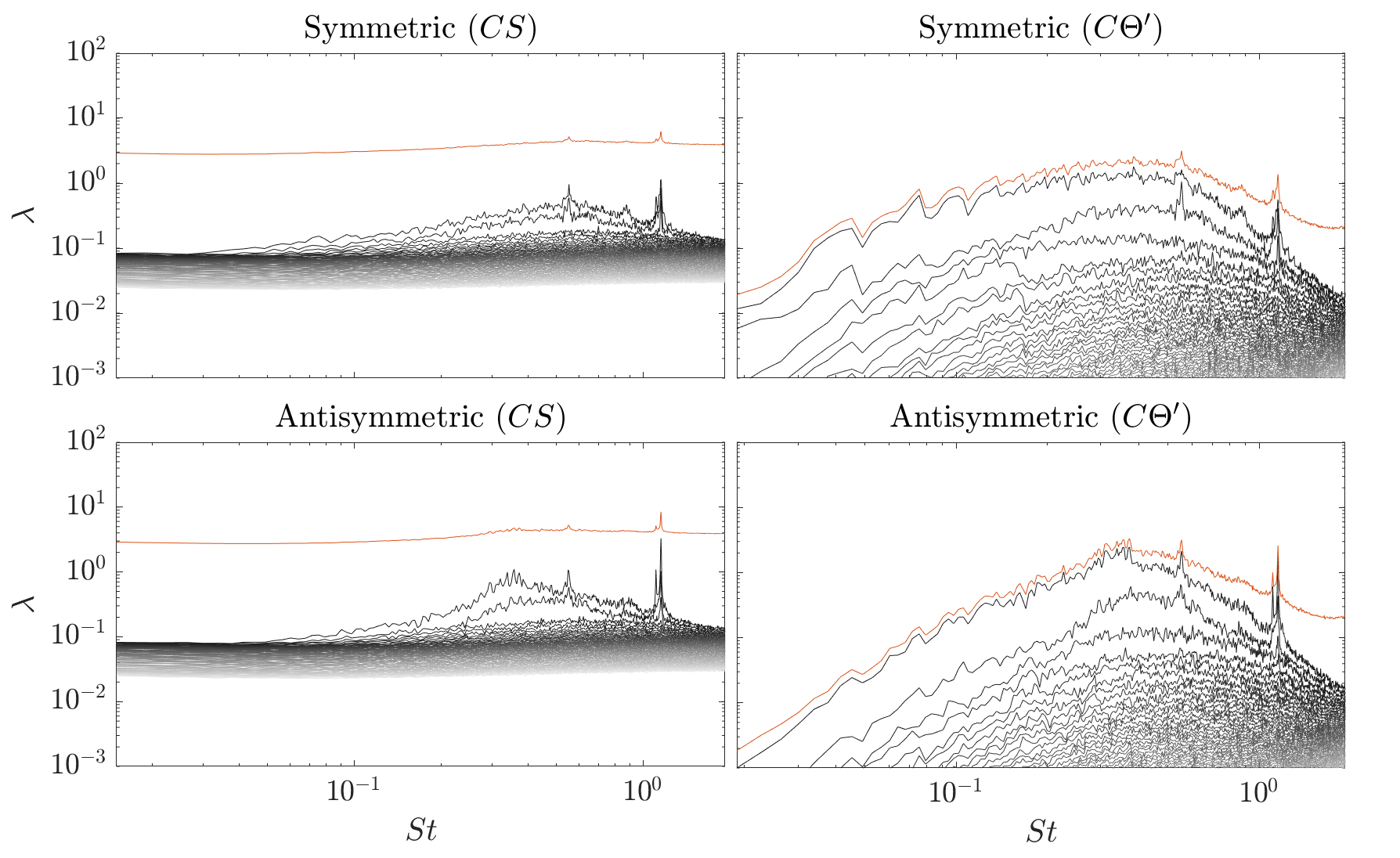}
\caption{Comparison of the SPOD spectra for $M_j = 1.54$. (left) $CS$ SPOD; (right) $C\Theta'$ SPOD; (top) symmetric case; (bottom) antisymmetric case. Each line corresponds to one SPOD mode. A grayscale is used to range from the most energetic mode to the least energetic one. The orange line represents the sum of the energy of all SPOD modes for each frequency.}
\label{fig:spectra_Mj1d54}
\end{figure}

\begin{figure}[t]
\centering
\includegraphics[width=0.99\textwidth]{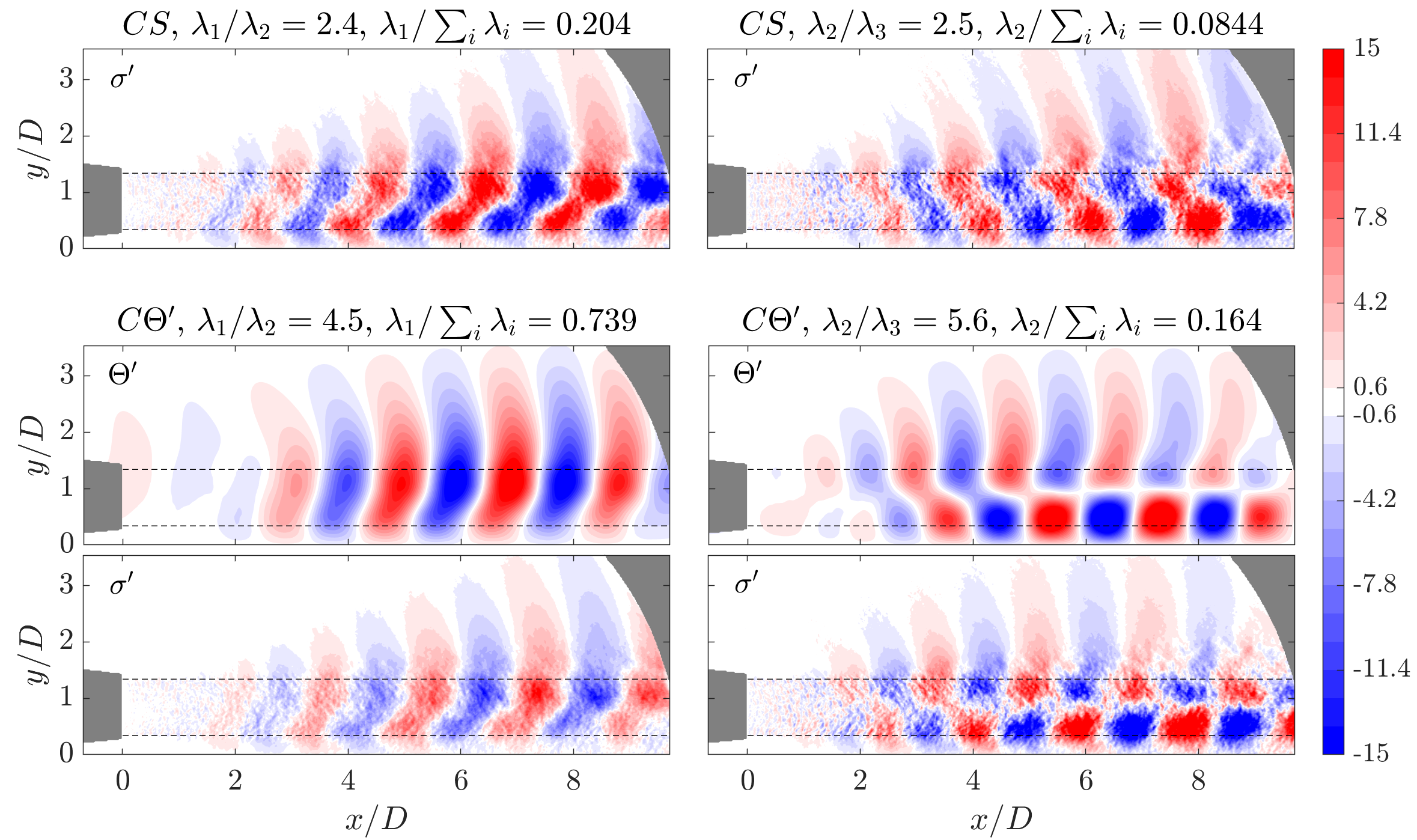}
\caption{Contours of the real part of the antisymmetric SPOD modes for $M_j = 1.54$ at $St = 0.37$: (top) $\sigma'$ field, $CS$ SPOD problem; (middle) $\Theta'$ field, $C\Theta'$ SPOD; (bottom) $\sigma'$ field, $C\Theta'$ SPOD. (left) mode 1; (right) mode 2. The grey regions represent the exterior nozzle surface and the part of the mirror support structure that is contained within the Schlieren window. The black dashed lines depict the inner and outer nozzle lip lines for each jet.}
\label{fig:SPOD1_ext_Mj1d54_St0d37_asym}
\end{figure}

\begin{figure}[t]
\centering
\includegraphics[width=0.99\textwidth]{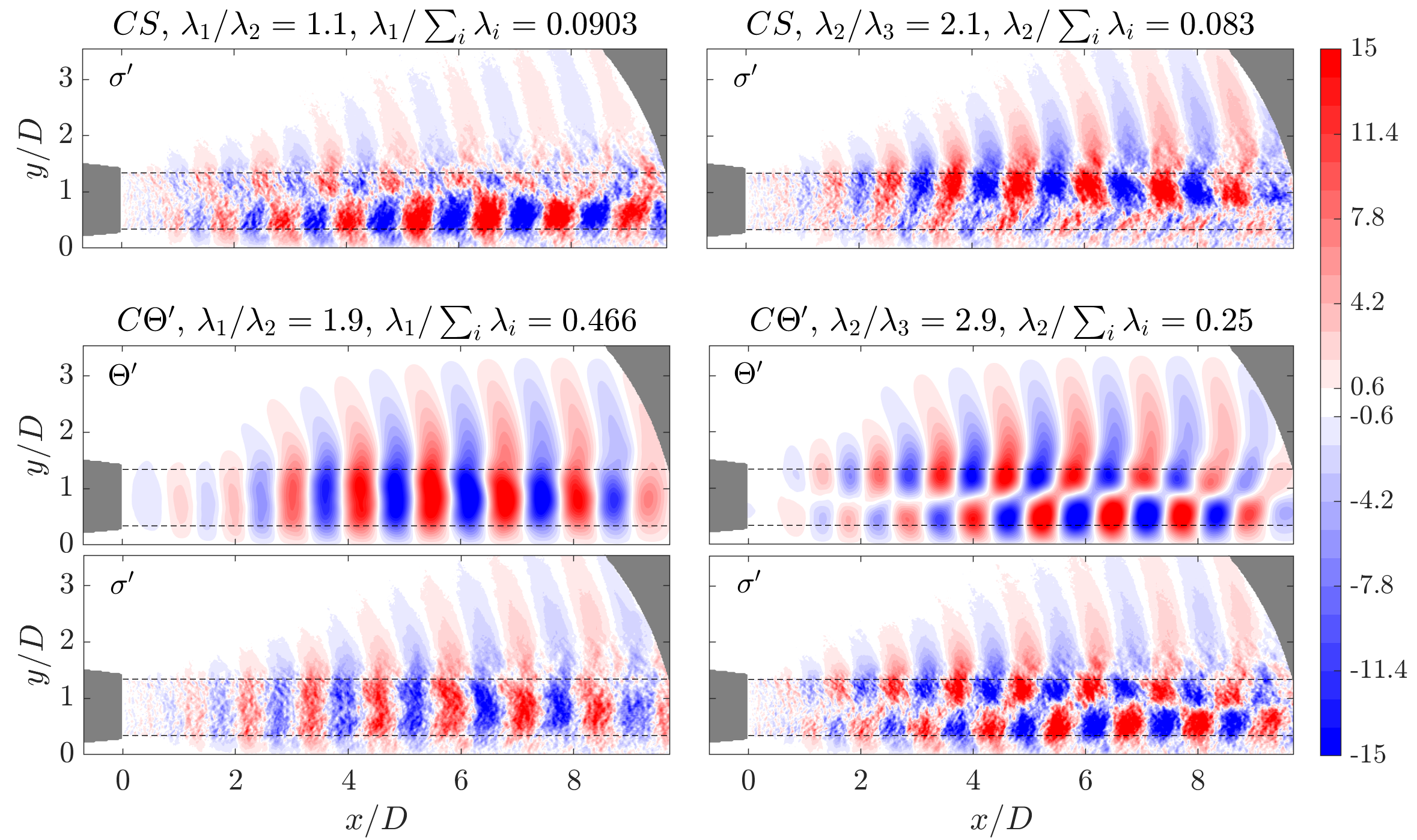}
\caption{Contours of the real part of the antisymmetric SPOD modes for $M_j = 1.54$ at $St = 0.61$: (top) $\sigma'$ field, $CS$ SPOD problem; (middle) $\Theta'$ field, $C\Theta'$ SPOD; (bottom) $\sigma'$ field, $C\Theta'$ SPOD; (left) mode 1; (right) mode 2.}
\label{fig:SPOD1_ext_Mj1d54_St0d61_asym}
\end{figure}

Perfectly-expanded jets do not present strong shock waves and their acoustic signature is characterized by broadband mixing-layer noise. In turn, Schlieren images are expected to be dominated by a large range of vortical fluctuations at all frequencies. SPOD based on the Schlieren images is nevertheless expected to recover the enhanced spatial coherence of the $\sigma'$ fluctuations in the mixing-noise frequency band, which would manifest as a bigger separation in the spectral density amplitude $\lambda_i$ of the leading SPOD modes.

Figure \ref{fig:spectra_Mj1d54} displays the symmetric and antisymmetric SPOD spectra obtained for $M_j = 1.54$ when using the correlations $CS$ and $C\Theta'$. The spectra are represented as a function of the Strouhal number, defined as $St = f D / u_j$, where $u_j$ is the isentropic jet velocity derived from $M_j$ and the isentropic nozzle exit temperature. For each frequency, the SPOD modes are ranked by their spectral density and are represented following a grayscale that ranges from the most relevant mode (mode 1, black) to the least relevant one (mode 57, white). In addition, a line representing the sum of the spectral density of all the SPOD modes ($\sum_i \lambda_i$) for each frequency is included in orange.

The SPOD spectra illustrate the expected broadband increase in spectral density ($St \approx 0.1$ to $1$) associated with the coherent wavepackets that grow in the shear layers of the jets, and which, for the current case, dominate for an antisymmetric coupling of the jets. Both the Schlieren and the $\Theta'$-based SPOD spectra show this trend, featuring the maximum spectral density at the same Strouhal numbers, $St = 0.3$ to $0.4$. Nevertheless, two important differences between both spectra stand out:

(i) As opposed to the results for $CS$, the spectral density summed for all modes for $C\Theta'$ (orange line) is substantially reduced except for the frequency range typical of mixing noise. This is consistent with the interpretation of $\Theta'$ as related to potential-energy fluctuations and their dominance for this range of frequencies. 

(ii) The $C\Theta'$ SPOD results exhibit a much lower-rank behavior than the Schlieren-based counterpart, as indicated by the small difference between the orange line and the line corresponding to the leading SPOD mode (mode 1). Therefore, most of the spectral density for $C\Theta'$ is accounted for by the leading SPOD mode. For increasing frequencies higher than $St=0.4$ the low-rank behavior is gradually lost, which could be a result of a physical loss of spatial coherence or an artifact of the finite pixel resolution.

The SPOD modes for both the $CS$ and the $C\Theta'$ SPOD solutions are illustrated in figures \ref{fig:SPOD1_ext_Mj1d54_St0d37_asym} and \ref{fig:SPOD1_ext_Mj1d54_St0d61_asym} for two different frequencies, respectively. Each figure shows the first (left column) and second (right column) SPOD modes in three different versions: (i) the Schlieren field obtained from the Schlieren SPOD ($CS$) (first row); (ii) the $\Theta'$ field obtained from the $\Theta'$ SPOD problem ($C\Theta'$) (second row); and (iii) the Schlieren field reconstructed from the $\Theta'$ SPOD problem ($C\Theta'$) as outlined in section \ref{sec:Sch_reconstruction} (third row). Due to the fact that the symmetric or antisymmetric fields are considered, only one of the two jets is shown in each contour plot. The quantity $\lambda_1/\lambda_2$ denotes the ratio between the energy of the first and the second SPOD modes, while $\lambda_1/ \sum_i \lambda_i$ is the portion of the total energy contained in the first mode (and similarly for $\lambda_2/\lambda_3$ and $\lambda_2/ \sum_i \lambda_i$).

Figure \ref{fig:SPOD1_ext_Mj1d54_St0d37_asym} shows the modes at the peak frequency of the broadband region for the antisymmetric coupling, located at $St = 0.37$. In agreement with previous investigations, the coherent structures resemble Kelvin-Helmholtz instability wavepackets that manifest as toroidal or flapping deformations of each jet column and present a noticeable downstream Mach wave radiation. Toroidal twin-jet modes are analogous to $m = 0$ modes in isolated round jets in the sense that they feature an axisymmetric amplitude function within each jet. Similarly, when a coupling exists between the oscillations of twin jets, it favors flapping motions over the $m=1$ helical modes typical of single round jets \cite{Rodriguez2022arXiv}. The nature of the Schlieren visualization prevents recovering oscillations that are anti-symmetric about the jet-containing plane; only symmetric (i.e. sinuous) or anti-symmetric (varicose) flapping motions can be observed. These oscillation modes are characterized by their vanishing amplitude along the axis of each jet.

However, owing to the mean flow interaction between both jets when they are closely spaced (such as in the configuration under study), the inner and outer shear layers of each jet have different thickness and shear magnitudes \cite{Goparaju2018}. As a result, the wavepackets no longer have a perfectly axisymmetric character, especially for lower frequencies which are associated with longer wavelengths, and the identification of the oscillation modes as $m = 0$ and $m = 1$ becomes less straightforward than for larger jet spacing~\cite{PadillaMontero2023}. In addition, the structures that characterize Schlieren-based SPOD modes in twin jets may contain a superposition of $m = 0$ and $m = 1$ waves in the same mode. This is the case of the Schlieren SPOD modes shown in the top row of figure \ref{fig:SPOD1_ext_Mj1d54_St0d37_asym}, where the supported coherent structures have an oblique orientation with respect to the jet axis and it is not evident whether their spatial structure corresponds to a toroidal or a flapping oscillation.

Looking at the second row of figure \ref{fig:SPOD1_ext_Mj1d54_St0d37_asym}, it can be observed that the structure of the $C\Theta'$ SPOD modes also captures the Kelvin-Helmholtz waves and their associated acoustic radiation. Moreover, it reveals that the $C\Theta'$ SPOD performs very well at separating the toroidal and the flapping structures into the first and second SPOD modes, respectively. This illustrates a key advantage of the methodology, namely that the $C\Theta'$ SPOD can provide a more robust extraction of coherent information in the system than that based on $CS$. Furthermore, note that $\lambda_1$ contains 74\% of the total spectral density of the $C\Theta'$ decomposition, while it only represents 20\% of the total for $CS$, which reflects the lower-rank behavior of the $C\Theta'$ decomposition described before.

The effectiveness of the $C\Theta'$ SPOD problem in separating and ranking the different coherent structures can then be brought to the Schlieren field by reconstructing the Schlieren modes corresponding to the $C\Theta'$ SPOD modes, as presented in the third row of figure \ref{fig:SPOD1_ext_Mj1d54_St0d37_asym}. The reconstructed Schlieren field for the second SPOD mode shows a much clearer $m = 1$ structure than the original Schlieren mode.

Figure \ref{fig:SPOD1_ext_Mj1d54_St0d61_asym} depicts the antisymmetric SPOD modes at a higher frequency ($St = 0.61$). In this case, it is more evident that $m = 0$ and $m = 1$ wavepackets are superposed in the original $CS$ SPOD modes. The $C\Theta'$ SPOD decomposition is found to effectively separate the structures into toroidal and flapping fluctuations in the first and second SPOD modes. The reconstructed Schlieren SPOD fields represent an improved description of the coherent structures associated with sound generation in the twin-jet system.

\subsection{Overexpanded condition}

\begin{figure}[t]
\centering
\includegraphics[width=0.99\textwidth]{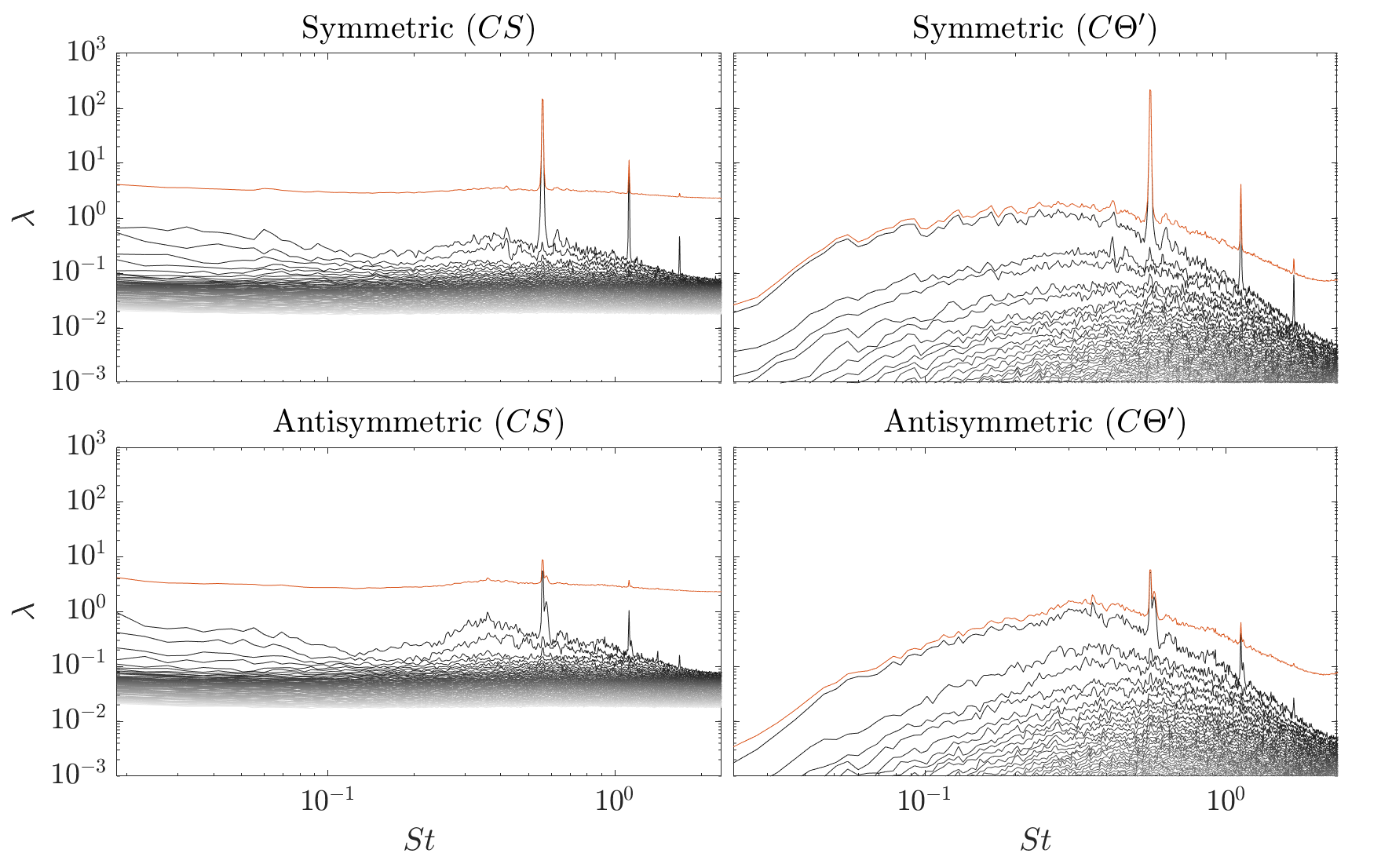}
\caption{Comparison of the SPOD spectra for $M_j = 1.26$. (left) $CS$ SPOD; (right) $C\Theta'$ SPOD; (top) symmetric case; (bottom) antisymmetric case.}
\label{fig:spectra_Mj1d26}
\end{figure}

\begin{figure}[t]
\centering
\includegraphics[width=0.55\textwidth]{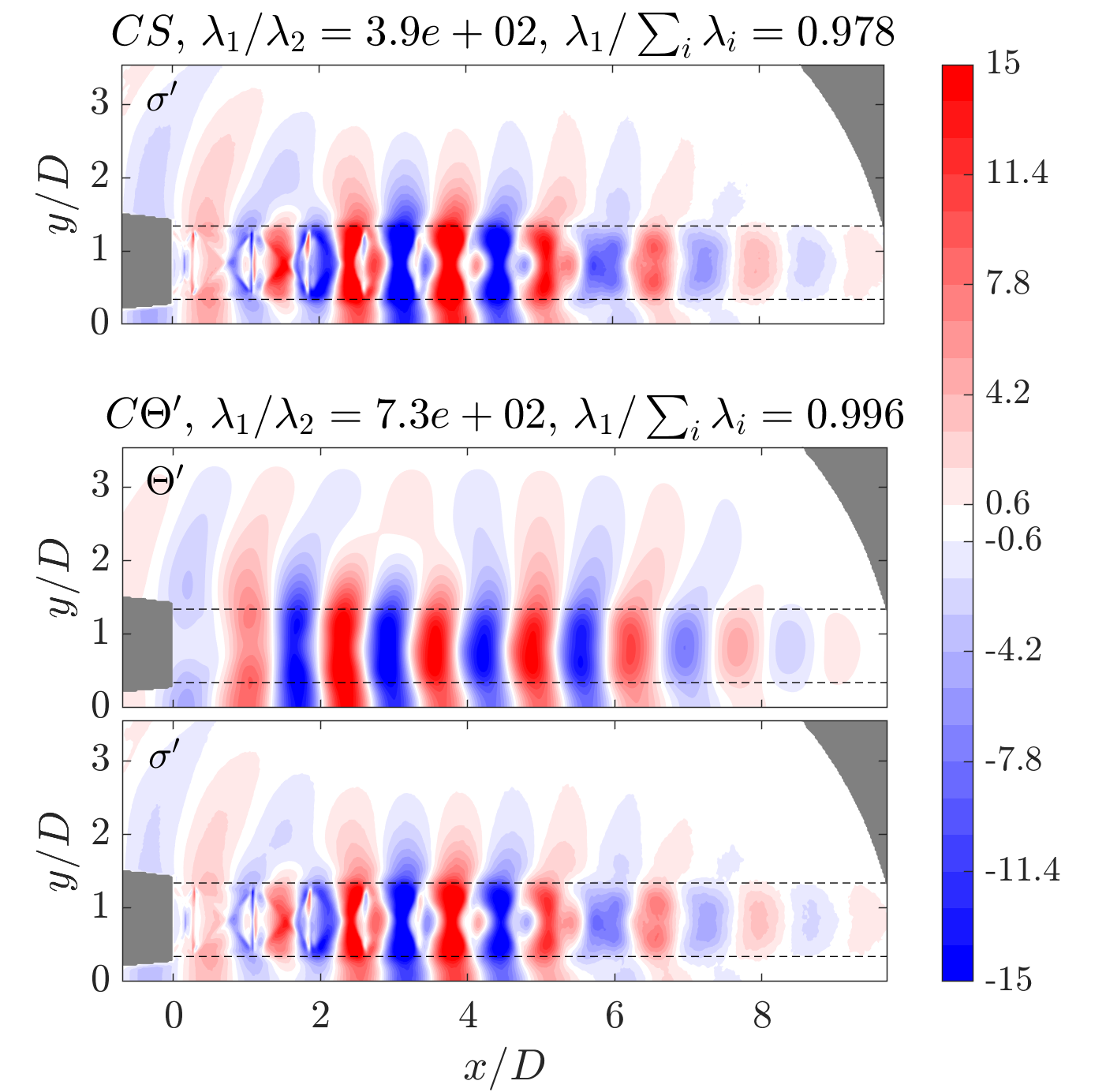}
\caption{Real part of the leading (mode 1) symmetric SPOD mode for $M_j = 1.26$ at $St = 0.56$, corresponding to the fundamental screech frequency: (top) $\sigma'$ field, $CS$ SPOD problem; (middle) $\Theta'$ field, $C\Theta'$ SPOD; (bottom) $\sigma'$ field, $C\Theta'$ SPOD.}
\label{fig:screech_mode_Mj1d26}
\end{figure}

\begin{figure}[t]
\centering
\includegraphics[width=0.99\textwidth]{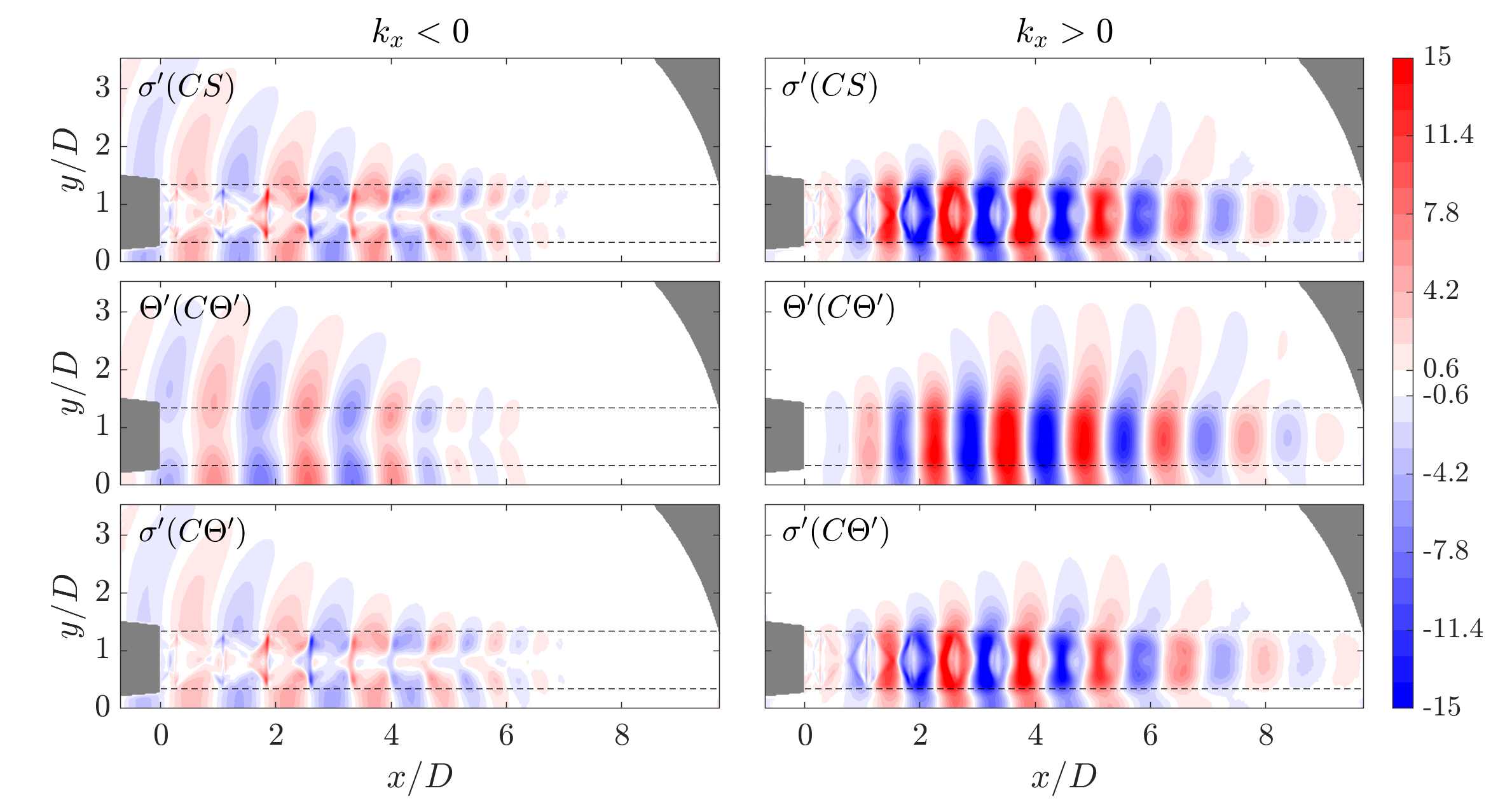}
\caption{Contours of the real part of the leading symmetric SPOD mode for $M_j = 1.26$ at $St = 0.56$: (left) $k_x < 0$ component; (right) $k_x > 0$ component; (top) $\sigma'$ field, $CS$ SPOD problem; (middle) $\Theta'$ field, $C\Theta'$ SPOD; (bottom) $\sigma'$ field, $C\Theta'$ SPOD.}
\label{fig:screech_mode_Mj1d26_kp_km}
\end{figure}

\begin{figure}[h!]
\centering
\includegraphics[width=0.99\textwidth]{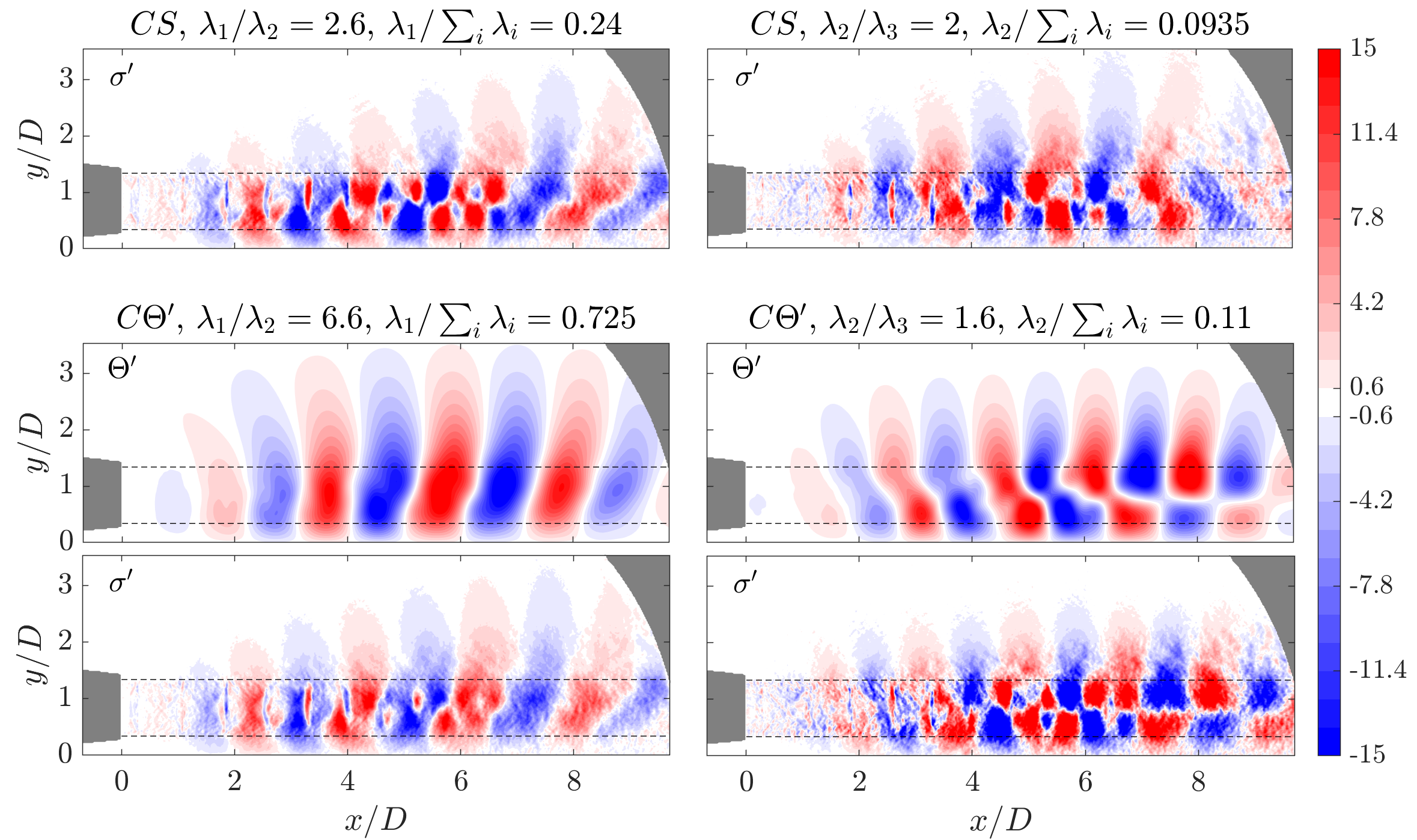}
\caption{Contours of the real part of the antisymmetric SPOD modes for $M_j = 1.26$ at $St = 0.36$: (top) $\sigma'$ field, $CS$ SPOD problem; (middle) $\Theta'$ field, $C\Theta'$ SPOD; (bottom) $\sigma'$ field, $C\Theta'$ SPOD; (left) mode 1; (right) mode 2.}
\label{fig:SPOD_ext_Mj1d26_St0d36_asym}
\end{figure}

Next, an isentropic Mach number at which the jets are significantly overexpanded is considered ($M_j = 1.26$). For this condition, shock cells are present in the core of the jets (see figure \ref{fig:Schlieren_Mj1d26}), and the spectra is expected to be dominated by the screech resonance mechanism (see \cite{Edgington-Mitchell2018}) together with mixing noise.

Figure~\ref{fig:spectra_Mj1d26} shows the symmetric and antisymmetric SPOD spectra for this case, obtained solving both the $CS$ and the $C\Theta'$ SPOD problems. The highest amplitude in the spectra is now encountered in the strong tonal peaks associated with the screech resonance. Three tonal peaks are visible, corresponding to the fundamental screech frequency ($St = 0.56$) and two of its harmonics. The tones are much more energetic for the symmetric spectra, indicating that the screech resonance for the condition under study favors a symmetric twin-jet coupling. Besides screech, the broadband signature associated with mixing noise can also be identified, once again showing higher spectral density for antisymmetric fluctuations. Except for the fundamental screech frequency, where both the $CS$ and the $C\Theta'$ spectra contain almost all of the energy in the first SPOD mode, the $C\Theta'$ SPOD features a much lower-rank behavior also for this $M_j$ along the entire broadband region.

\begin{figure}[h!]
\centering
\includegraphics[width=0.99\textwidth]{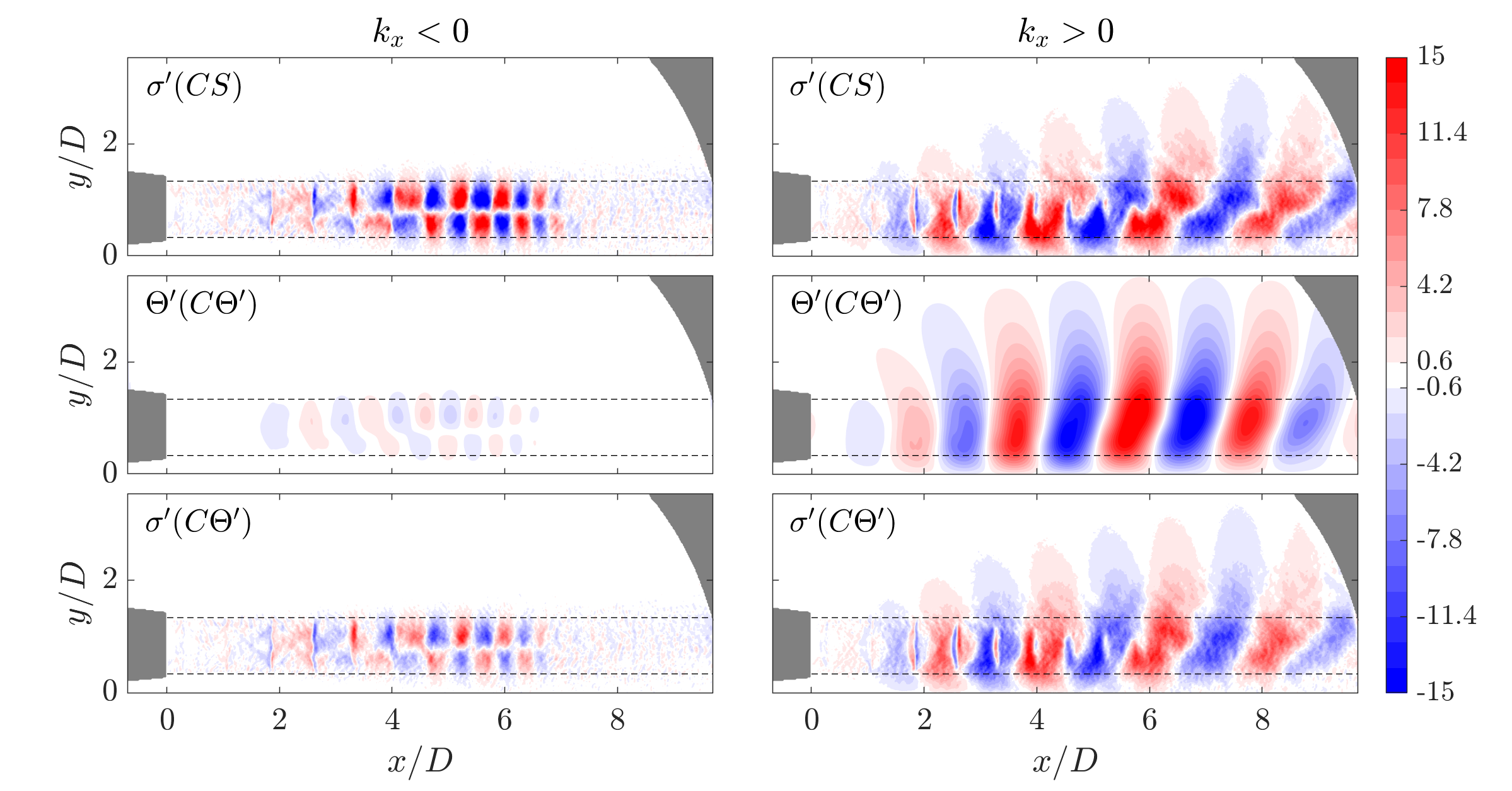}
\caption{Contours of the real part of the antisymmetric SPOD mode 1 for $M_j = 1.26$ at $St = 0.36$: (left) $k_x<0$ component; (right) $k_x>0$ component; (top) $\sigma'$ field, $CS$ SPOD problem; (middle) $\Theta'$ field, $C\Theta'$ SPOD; (bottom) $\sigma'$ field, $C\Theta'$ SPOD.}
\label{fig:SPOD1_ext_Mj1d26_St0d36_asym_kp_km}
\end{figure}

\begin{figure}[h!]
\centering
\includegraphics[width=0.99\textwidth]{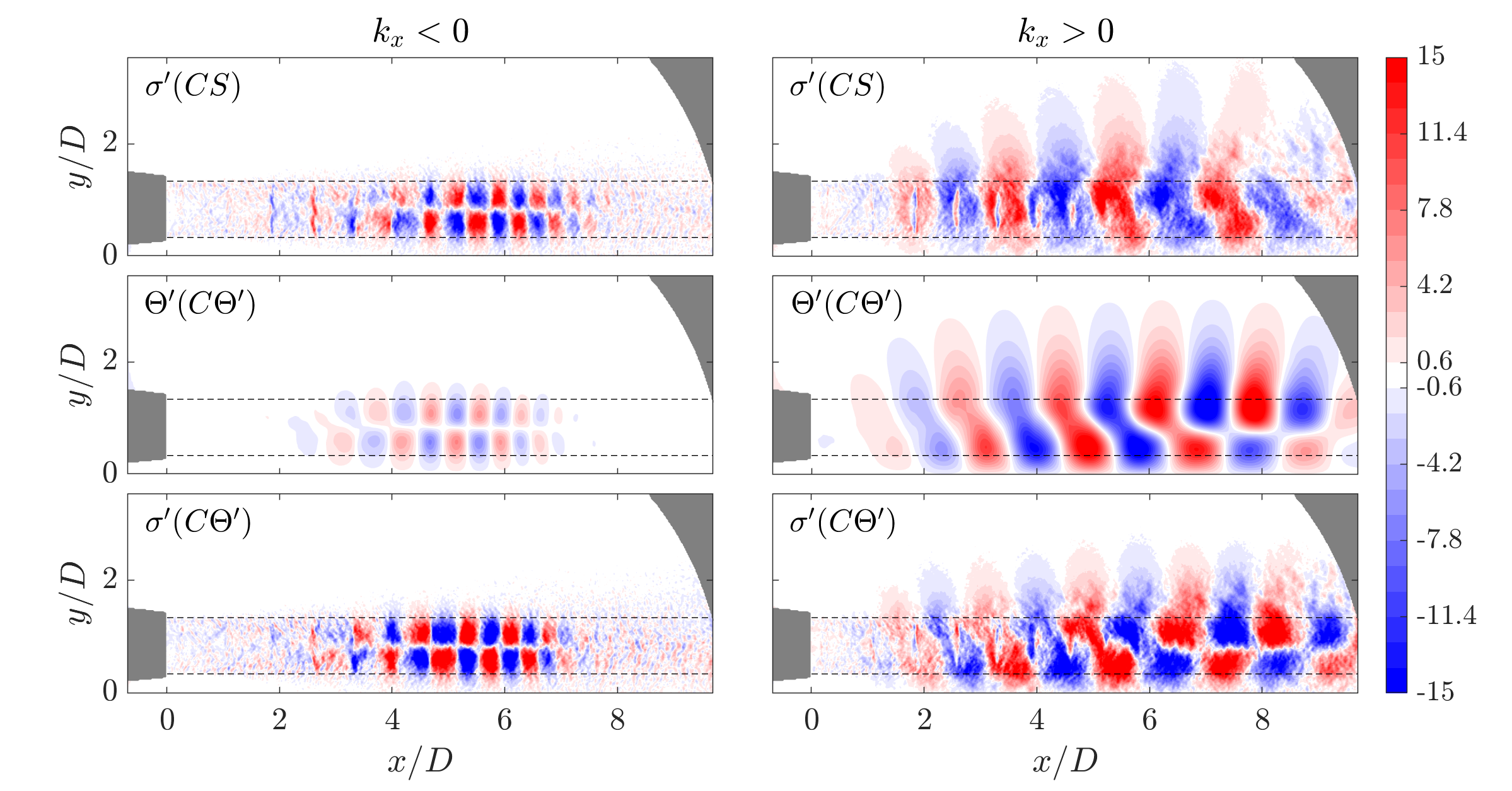}
\caption{Contours of the real part of the antisymmetric SPOD mode 2 for $M_j = 1.26$ at $St = 0.36$: (left) $k_x<0$ component; (right) $k_x>0$ component; (top) $\sigma'$ field, $CS$ SPOD problem; (middle) $\Theta'$ field, $C\Theta'$ SPOD; (bottom) $\sigma'$ field, $C\Theta'$ SPOD.}
\label{fig:SPOD2_ext_Mj1d26_St0d36_asym_kp_km}
\end{figure}

The SPOD modes for the over-expanded jet condition are presented in figures \ref{fig:screech_mode_Mj1d26} to \ref{fig:SPOD2_ext_Mj1d26_St0d36_asym_kp_km}. First, the symmetric first SPOD mode corresponding to the fundamental screech frequency ($St = 0.56$) is analyzed, illustrated in figure \ref{fig:screech_mode_Mj1d26}. As in the previous section, three different fields are represented: the Schlieren field corresponding to $CS$, the $\Theta'$ field corresponding to $C\Theta'$, and the Schlieren field corresponding to $C\Theta'$.

The $CS$ SPOD mode shows structures that correspond to toroidal wavepackets modulated by the shock cells, together with downstream and upstream acoustic radiation. Due to the strong resonance that originates the screech, in this case there is no difficulty in identifying the $m = 0$ structure in the amplitude function ($\lambda_1$ represents nearly 98\% of the total energy). The $C\Theta'$ SPOD mode structure also contains most of these features. Nonetheless, note that no signature of the shock-cell modulation is visible in the $\Theta'$ field, which illustrates how, within the frame of momentum potential theory, shock cells do not contribute directly to the irrotational fluctuation components.

In order to separate the upstream- and downstream-propagating wave components present in the SPOD mode associated to the fundamental screech frequency, the spatial DFT of the SPOD mode is calculated along $x$ to obtain a distribution of the mode energy as a function of the streamwise wavenumber $k_x$. This enables the separation of those energetic components with $k_x > 0$, i.e. positive phase speed, from those components with $k_x < 0$, i.e. negative phase speed. The spatial Fourier transform of the SPOD mode is then inverted by keeping only the $k_x > 0$ or the $k_x < 0$ components, resulting in the structures represented in figure \ref{fig:screech_mode_Mj1d26_kp_km}. The right column shows waves that propagate downstream within the screech mode, consisting of a Kelvin-Helmholtz wavepacket and its associated acoustic radiation. The left column contains mostly waves that propagate upstream, and which take the form of acoustic waves with spatial support that extends into the freestream, the jet shear layers and in the subsonic regions of the jet core. It is important to note that the actual direction of propagation of energy in the flow is dictated by the sign of the group velocity ($c_g = \partial \omega / \partial k_x$), rather than that of the phase velocity. Although most waves with negative phase speed also feature negative group velocity and thus propagate upstream (and vice-versa), there are also certain waves with negative phase speed that have a positive group velocity and move downstream, such as acoustic waves that travel within the supersonic jet core region. The $C\Theta'$ SPOD solution also contains the energetic signature of both types of waves, and consequently the reconstructed parts of the Schlieren mode look identical to the original one.

The second frequency analyzed for this operating condition corresponds to the maximum spectral amplitude in the mixing-noise broadband region, namely, $St = 0.36$, which is higher for the antisymmetric case. Figure \ref{fig:SPOD_ext_Mj1d26_St0d36_asym} depicts the respective SPOD modes for this frequency. Similarly to the previous section, both the first (left column) and second (right column) SPOD modes are shown. In this case, the $CS$ SPOD modes show apparently a superposition of three structures with two different wavenumbers: one of them non-radiating and mainly confined within the jet core, featuring the larger wavenumber, and the other two consisting of toroidal and flapping Kelvin-Helmholtz signatures of identical wavelength and their associated downstream acoustic radiation, similarly to the modes discussed in the previous section.

The separation into $k_x < 0$ and $k_x > 0$ components, crucial in this case to understand the structure of the modes, is shown in figures \ref{fig:SPOD1_ext_Mj1d26_St0d36_asym_kp_km} and \ref{fig:SPOD2_ext_Mj1d26_St0d36_asym_kp_km} for the first and second SPOD modes, respectively. The splitting reveals a positive phase-speed component consisting of mixed $m = 0$ and $m = 1$ wavepacket structures, together with a negative phase-speed component formed by an $m = 1$ trapped acoustic wave with support within the supersonic jet-core region. Such acoustic wave, which strongly resembles the so-called duct modes~\cite{Towne2017,Edgington-Mitchell2021}, is in this case an example of a downstream-propagating wave with negative phase velocity. Its positive group velocity can be estimated by performing the DFT along $x$ and tracking the evolution of its dominant streamwise wavenumber as a function of frequency, as illustrated in the $St$-$k_x$ maps found in \cite{PadillaMontero2023b}. Note that since the core of the present jets is mostly supersonic, such a wave can only propagate energy downstream.

The second and third rows of figures \ref{fig:SPOD_ext_Mj1d26_St0d36_asym} to \ref{fig:SPOD2_ext_Mj1d26_St0d36_asym_kp_km} show the effectiveness of $C\Theta'$ SPOD in separating the different coherent structures found at this frequency. The first $C\Theta'$ SPOD mode comprises the toroidal wavepacket structure, together with a weak signature of the trapped acoustic wave. In turn, the second $C\Theta'$ SPOD mode consists of the flapping wavepacket together with a much stronger signature of the $m = 1$ acoustic wave. These observations are further reflected on the reconstructed Schlieren fields of the $C\Theta'$ SPOD modes. Whereas the $CS$ SPOD superposes all three $m = 0$ and $m = 1$ waves in the first and second modes, the $C\Theta'$ SPOD  tends to separate them effectively into different SPOD modes. The $m = 0$ wave is recovered in the first SPOD mode, while the two $m = 1$ waves are found in the second SPOD mode, and are easily separated by the phase-speed splitting. The latter behavior is more advantageous for the visualization and physical interpretation of the different type of oscillations undergone by the jets.

\section{Conclusions} \label{sec:conclusions}

A methodology to improve the extraction of coherent information from high-speed Schlieren images in supersonic twin jets has been presented. Following Prasad \& Gaitonde~\cite{Prasad2022}, the proposed approach makes use of the momentum decomposition introduced by Doak~\cite{Doak1989} and originally applied by Jordan et al.~\cite{Jordan2013}, to derive a Poisson equation relating the Schlieren fluctuation field with the streamwise gradient of the momentum potential fluctuations integrated along the line of sight, hereby denoted by $\Theta'$. As opposed to Schlieren images that are dominated in the turbulent mixing region by vortical fluctuations of a broad range of length scales, $\Theta'$ provides a representation of the potential (acoustic and thermal) energy embedded in the fluctuations, and its spatial structure in the frequency domain is remarkably coherent. By integrating $\Theta'$ within the spectral proper orthogonal decomposition (SPOD) framework, coherent structures of the twin-jet field are obtained which follow a much lower-rank behavior than those obtained when the SPOD problem is based on Schlieren images alone.

The SPOD modes calculated when $\Theta'$ is used to construct the cross-spectral density tensor retain the coherent structure and the acoustic radiation associated with Kelvin-Helmholtz wavepackets and the screech resonance mechanism, but are substantially more effective in separating the toroidal ($m = 0$) and flapping ($m = 1$) fluctuations coexisting in coupled twin jets, which are recovered as independent SPOD modes. In addition, as the modal decomposition is based on a variable which represents potential-energy fluctuations, the extracted coherent structures are more closely related to the sound-generating mechanisms of the system and facilitate a more detailed analysis of the underlying physics.

\begin{acknowledgements}
The work of D.R. and I.P. is partially funded by the Government of the Community of Madrid within the multi-annual agreement with Universidad Polit\'ecnica de Madrid through the Program of Excellence in Faculty (V-PRICIT line 3) and the Program of Impulse of Young Researchers (V-PRICIT lines 1 and 3, Grant No. APOYO-JOVENES-WYOWRI-135-DZBLJU). I.P. also thanks financial support from the European Union's NextGenerationEU fund (RD 289/2021: Ayudas Margarita Salas para la formaci\'{o}n de j\'{o}venes doctores). The authors also acknowledge St\`{e}ve Girard and Damien Eysseric for the setup, technical assistance and critical assessment of the Schlieren measurements at the PROM\'{E}T\'{E}E platform of Institut Pprime.
\end{acknowledgements}

\section*{Declarations}

\section*{Ethical Approval}

Not applicable.

\section*{Funding}

Partial financial support was received from the Government of the Community of Madrid within the multi-annual agreement with Universidad Polit\'ecnica de Madrid through the Program of Excellence in Faculty (V-PRICIT line 3) and the Program of Impulse of Young Researchers (V-PRICIT lines 1 and 3, Grant No. APOYO-JOVENES-WYOWRI-135-DZBLJU). Partial funding was also received from the European Union's NextGenerationEU program (RD 289/2021: Ayudas Margarita Salas para la formaci\'{o}n de j\'{o}venes doctores).

\section*{Availability of data and materials}

The data sets generated and/or analyzed in this investigation are available from the corresponding author upon reasonable request.

\begin{appendices}

\section{Unphysical harmonic waves in $\Theta'$ and the impact of the filtering process used to remove them} \label{sec:AppA}

As discussed in \S \ref{sec:Poisson_solution}, the solution of the Poisson equation \eqref{eq:Poisson_sch} to compute $\Theta'$ fields from Schlieren fluctuations would require impractically large domains to reach boundaries at which the flow is irrotational, or appropriate boundary conditions for $\Theta'$. Due to the fact that, in general, it is not possible to determine the boundary values that would provide an irrotational potential field at the boundaries of the domain, an unphysical harmonic component is introduced in the solution when other boundary conditions (such as homogeneous Dirichlet or periodic conditions) are enforced. The solution of the Poisson problem in the spectral domain allows the identification of such harmonic component as spurious waves in the solution, and to filter them out based on their unphysical phase velocities.

\begin{figure}[t]
\centering
\includegraphics[width=0.99\textwidth]{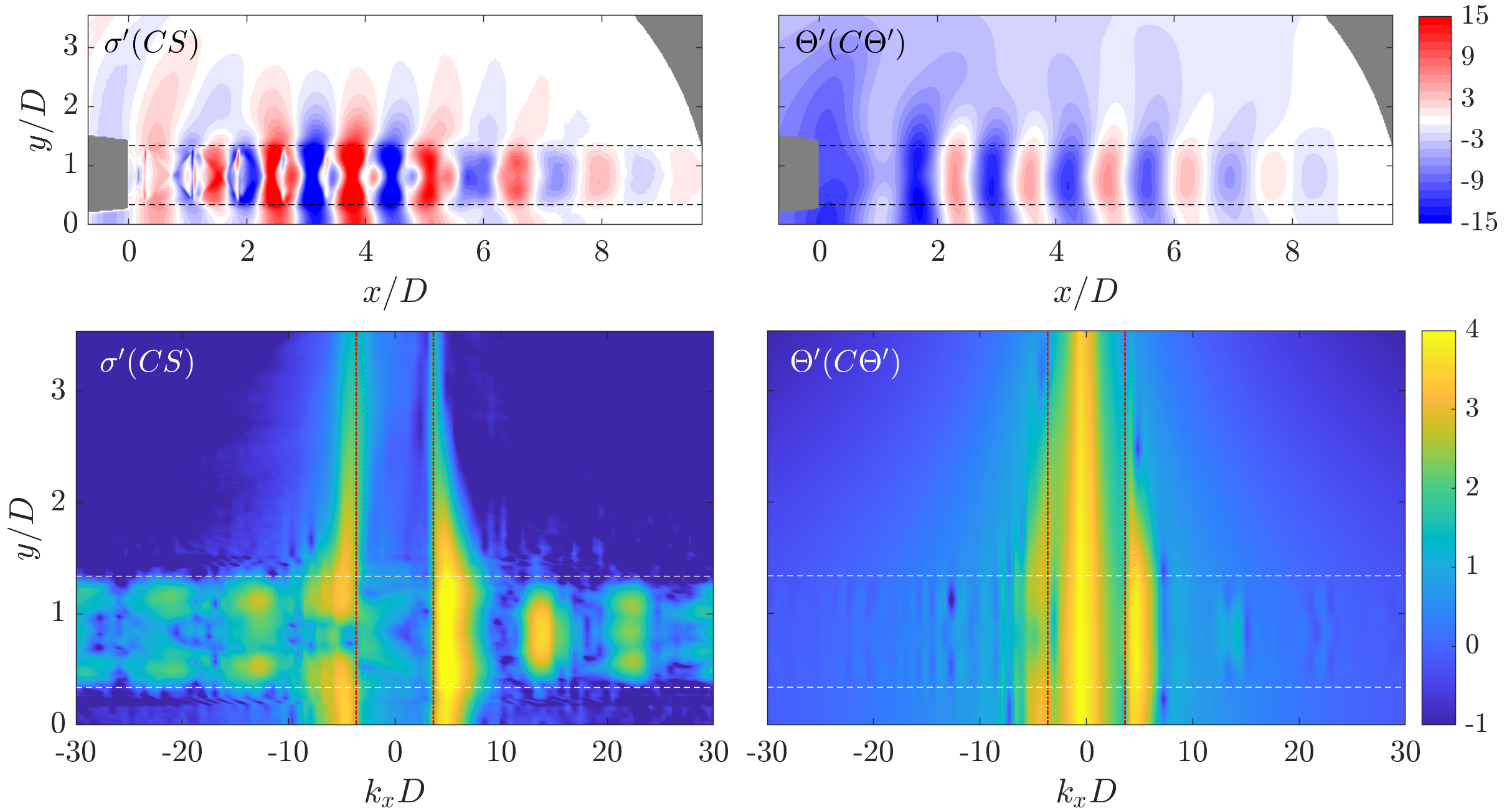}
\caption{Symmetric SPOD mode 1 for $M_j = 1.26$ at $St = 0.56$: (left) results from the Schlieren-based SPOD problem; (right) results from the $\Theta'$-based SPOD problem without filtering unphysical harmonic waves; (top) contours of the real part of the mode amplitude function; (bottom) Fourier transform of the SPOD mode along $x$, represented as contours of the logarithm of the energy magnitude. The red dash-dot lines denote the acoustic wavenumber associated to the frequency under investigation, given by the value at which the phase speed equals the freestream speed of sound. The white dashes lines represent the nozzle lip lines.}
\label{fig:AppA_1}
\end{figure}

To illustrate the impact of the unphysical harmonic waves on the calculated SPOD modes when they are not removed, figure \ref{fig:AppA_1} depicts the amplitude function of the first symmetric SPOD mode obtained for $M_j = 1.26$ at the fundamental screech frequency ($St = 0.56$). The left column shows the Schlieren SPOD mode obtained from the Schlieren-based SPOD problem, originally presented in figure \ref{fig:screech_mode_Mj1d26} but repeated here for convenience, while the right column shows the $\Theta'$ SPOD mode obtained without filtering the unphysical energetic component. In contrast to the $\Theta'$ amplitude function shown in figure \ref{fig:screech_mode_Mj1d26}, when the energetic harmonic waves are not filtered from $\tilde{\Theta}'$, the $\Theta'$ SPOD mode is contaminated with small wavenumber (large wavelength) unphysical structures, as shown in the top right contour plot of figure \ref{fig:AppA_1}. The energetic signature of these waves can be easily identified by performing the Fourier transform of the SPOD mode along $x$, as shown in the bottom right plot of the figure. For comparison, the energy distribution of the original Schlieren mode is also displayed in the bottom left picture. As expected, the energy signature of this mode mainly consists of the Kelvin-Helmholtz energy band, located at positive $k_x$ slightly larger than the acoustic wavenumber, and an acoustic energy band resulting from the upstream- and downstream-traveling acoustic waves excited in the screech resonance, mainly located at negative wavenumbers below the acoustic one.

\begin{figure}[t]
\centering
\includegraphics[width=0.99\textwidth]{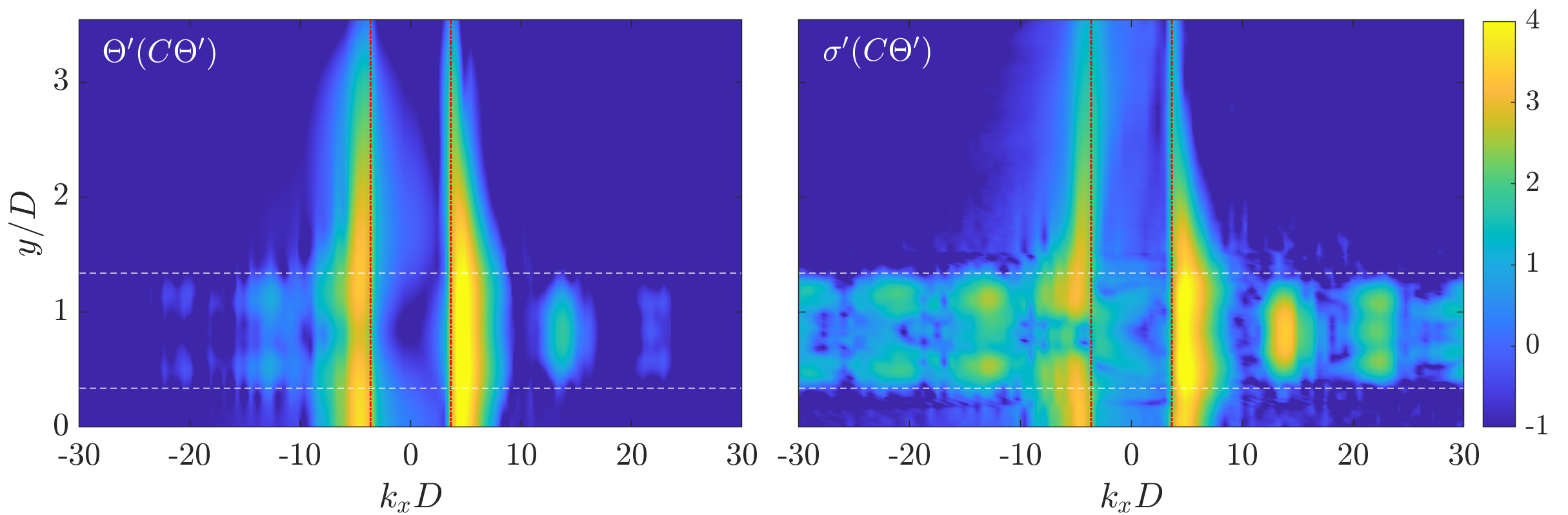}
\caption{Contours of the streamwise Fourier transform of the first symmetric SPOD mode for $M_j = 1.26$ at $St = 0.56$, obtained from the $\Theta'$-based SPOD problem with filtered unphysical harmonic waves (see figure \ref{fig:screech_mode_Mj1d26}): (left) logarithm of the energy magnitude of the $\Theta'$ SPOD mode; (right) logarithm of the energy magnitude of the Schlieren mode reconstructed from the $\Theta'$-based SPOD solution.}
\label{fig:AppA_2}
\end{figure}

By comparison with the energy distribution of the original Schlieren mode, the unfiltered $\Theta'$ mode contains an additional strong energetic signature centered at $k_x = 0$, which is not visible in the Schlieren map. This energetic band is purely due to the artificial harmonic waves introduced in the numerical solution of the Poisson equation. Note that its signature can be distinguished from the physical energy components described above, and that it is composed of structures that travel at unphysical, high-supersonic speeds. This result explains the origin of the filtering procedure described in \S \ref{sec:Poisson_solution}.

By removing the energetic components with wavenumbers below a chosen phase-speed threshold, in this case $c_{ph,c} = 1.2 c_\infty$, the screech SPOD modes based on $C\Theta'$ presented in figure \ref{fig:screech_mode_Mj1d26} are obtained. The energy signature of those modes is shown in figure \ref{fig:AppA_2}. Note that, now, the energy of the $\Theta'$ mode only shows the signatures of the physical waves playing a role in the screech resonance. In addition, the reconstructed Schlieren mode has an energy distribution that is practically identical to the original Schlieren SPOD mode, which demonstrates the effectiveness of the filtering approach and that it does not impact the energy of the physical information of interest.

\begin{figure}[t]
\centering
\includegraphics[width=0.99\textwidth]{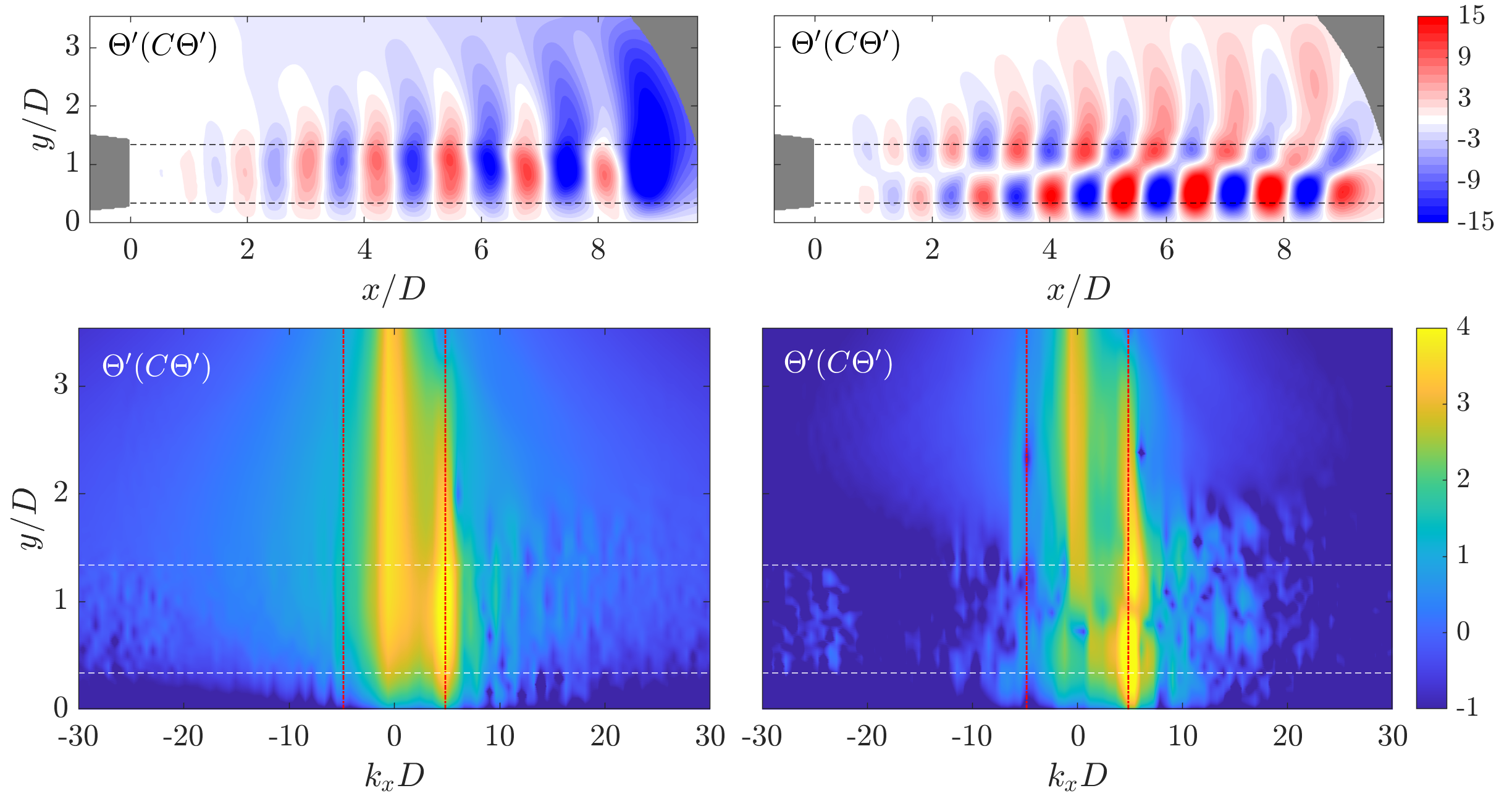}
\caption{Antisymmetric SPOD modes for $M_j = 1.54$ at $St = 0.61$ obtained from the $\Theta'$-based SPOD calculation without filtering unphysical harmonic waves: (left) first SPOD mode; (right) second SPOD mode; (top) contours of the real part of the mode amplitude function; (bottom) Fourier transform of the SPOD mode along $x$, represented as contours of the logarithm of the energy magnitude.}
\label{fig:AppA_3}
\end{figure}

\begin{figure}[h!]
\centering
\includegraphics[width=0.99\textwidth]{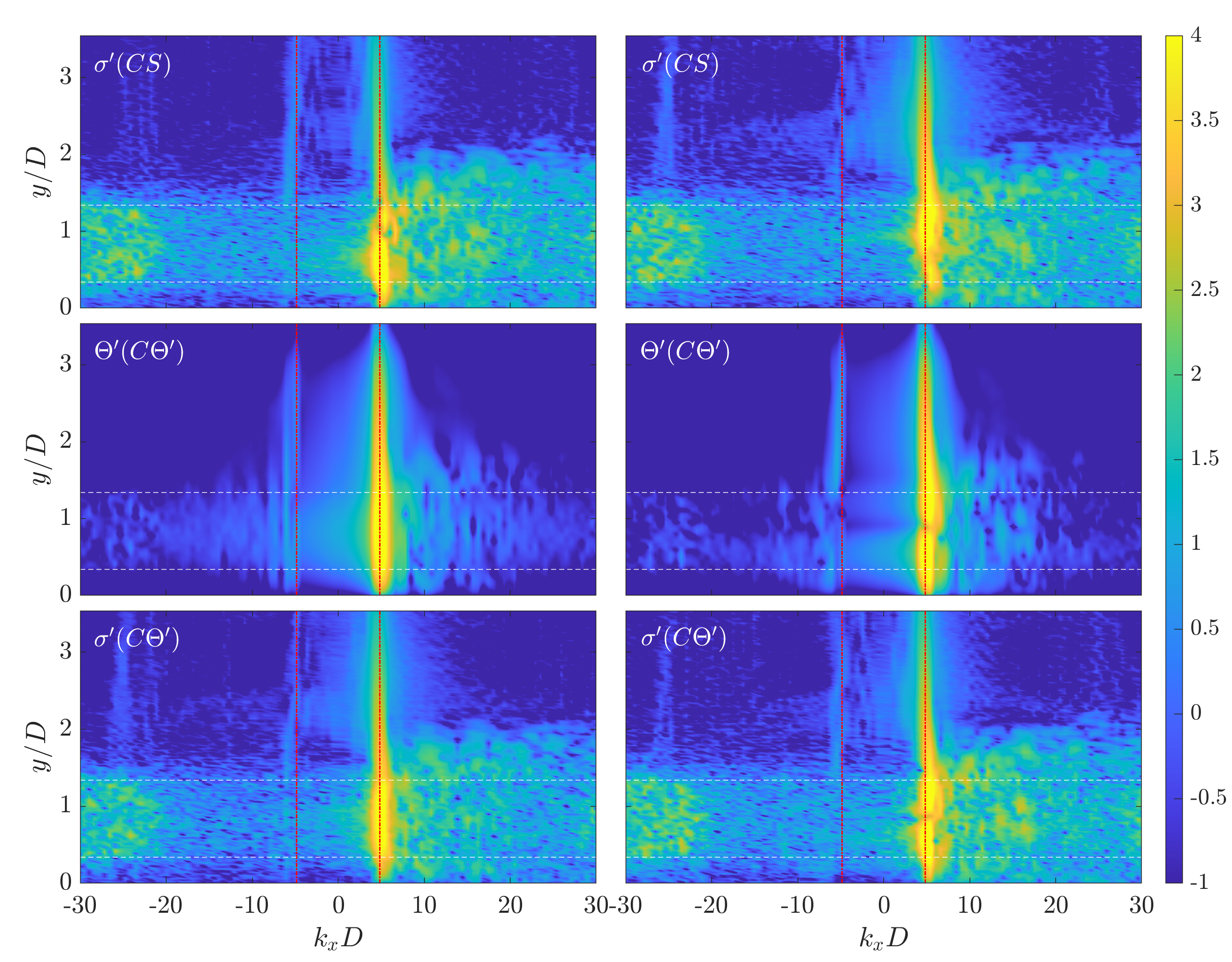}
\caption{Contours of the streamwise Fourier transform of the symmetric SPOD modes for $M_j = 1.54$ at $St = 0.61$, presented as the logarithm of the energy magnitude: (left) first SPOD mode; (right) second SPOD mode; (top) Schlieren SPOD mode obtained from the Schlieren-based SPOD problem; (middle) $\Theta'$ SPOD mode obtained from the $\Theta'$-based SPOD problem with filtered unphysical harmonic waves; (bottom) logarithm of the energy magnitude of the Schlieren mode reconstructed from the $\Theta'$-based SPOD solution with filtered unphysical harmonic waves. The contour plots in this figure have a one-to-one correspondence with the amplitude functions displayed in figure \ref{fig:SPOD1_ext_Mj1d54_St0d61_asym}.}
\label{fig:AppA_4}
\end{figure}

Another case is also reported to justify the need to identify and control the unphysical harmonic waves introduced by the \textit{ad hoc} imposition of boundary conditions in the Poisson equation. Figure \ref{fig:AppA_3} shows the unfiltered antisymmetric $\Theta'$ SPOD results for $M_j = 1.54$ at $St = 0.61$, in direct comparison with the filtered results already presented in figure \ref{fig:SPOD1_ext_Mj1d54_St0d61_asym}. Both the first and second SPOD modes are depicted in the top row, once again showing the impact of the unphysical waves on the amplitude functions, which is more significant for the first mode in this case. This is corroborated by the distribution of the energy magnitude over the streamwise wavenumbers, illustrated in the bottom row. Two main energy bands are visible for each mode, the one associated with the Kelvin-Helmholtz waves, located very close to the acoustic wavenumber, and the one associated with the harmonic unphysical energy components, which is stronger for the first mode. Note that in this case no physical energy components are present at negative wavenumbers, which reflects the fact that the mixing noise associated with Kelvin-Helmholtz instabilities is the dominant mechanism in the perfectly-expanded regime.

The corresponding energy signatures for the filtered modes are shown in figure \ref{fig:AppA_4} for comparison, including the Schlieren modes from the $CS$ SPOD and the $C\Theta'$ SPOD results. The removal of the unphysical harmonic waves from the $C\Theta'$ SPOD problem yields modes whose energetic structure is perfectly consistent with that of the original Schlieren modes, providing evidence that the applied filtering procedure is suitable for the problem under study.

\subsection{Effect of the phase-speed threshold employed for filtering the most energetic unphysical harmonic waves}

\begin{figure}[!ht]
\centering
\includegraphics[width=0.99\textwidth]{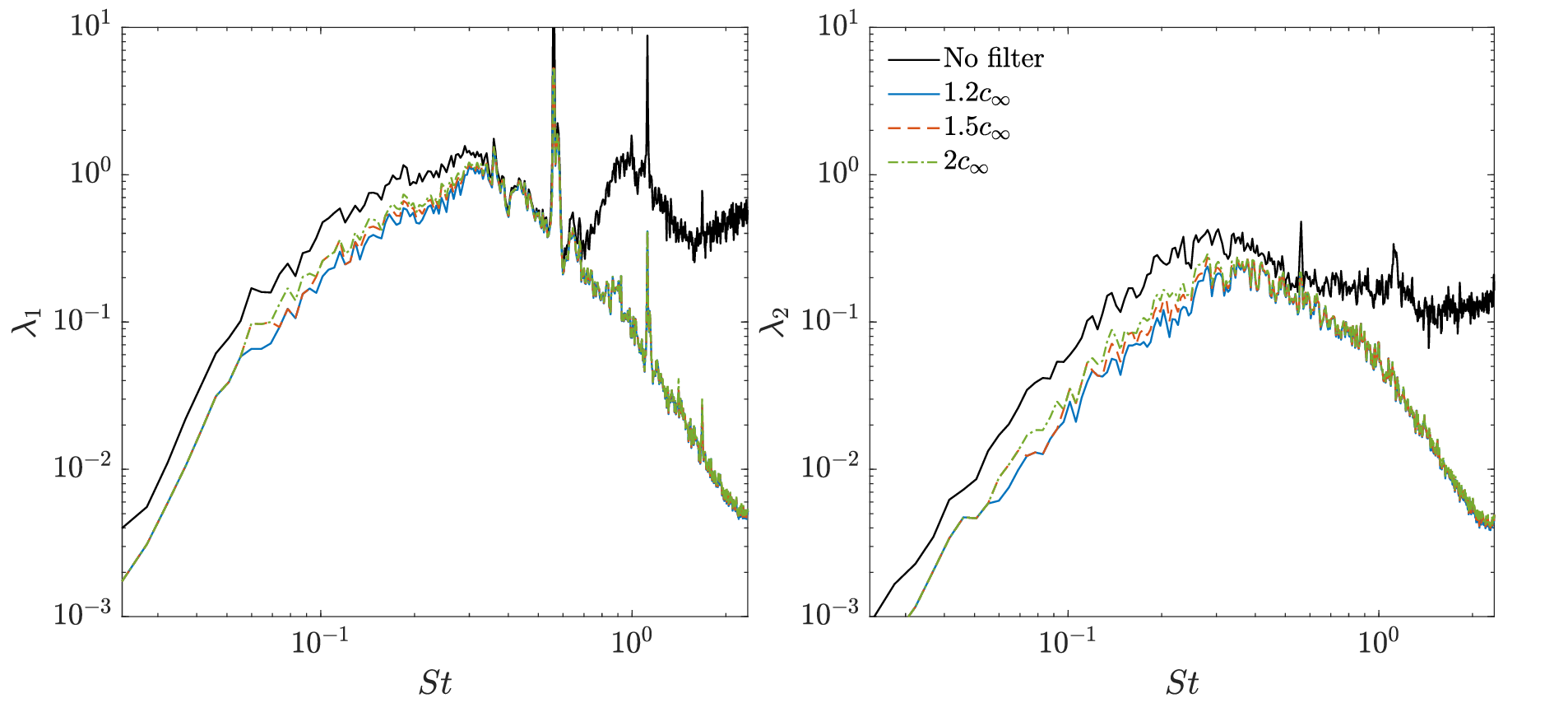}
\caption{Comparison of antisymmetric $\Theta'$ SPOD spectra for $M_j = 1.26$ using different phase-speed thresholds for removing unphysical harmonic waves: (left) first SPOD mode; (right) second SPOD mode. The unfiltered case is also added for comparison.}
\label{fig:AppA_5}
\end{figure}

\begin{figure}[t]
\centering
\includegraphics[width=0.99\textwidth]{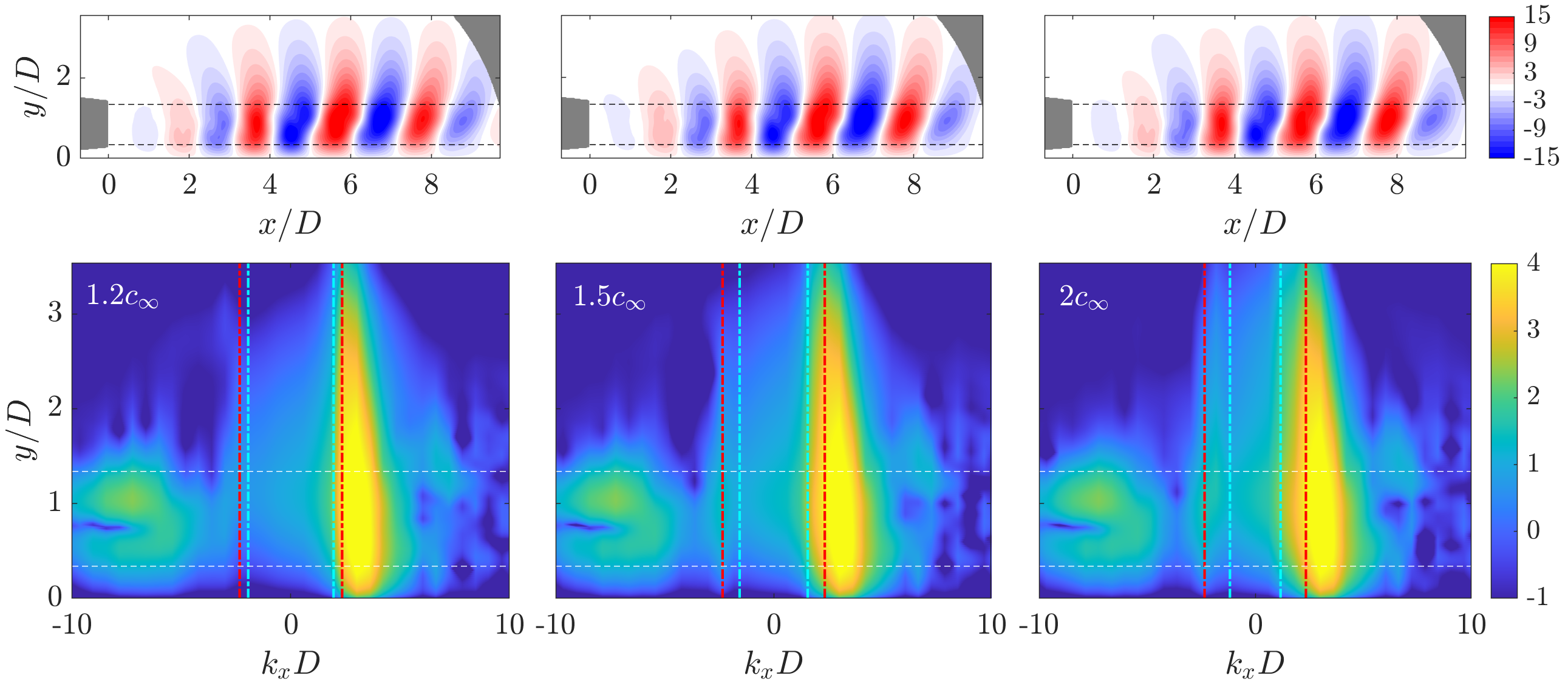}
\caption{Comparison of the first antisymmetric $\Theta'$ SPOD mode for $M_j = 1.26$, $St = 0.36$ using different phase-speed thresholds for removing unphysical harmonic waves: (left) $c_{ph,c} = 1.2 c_\infty$; (middle) $c_{ph,c} = 1.5 c_\infty$; (right) $c_{ph,c} = 2 c_\infty$; (top) real part of the SPOD mode amplitude function; (bottom) logarithm of the energy magnitude from the streamwise Fourier transform of the SPOD mode. The light blue dash-dot lines denote the wavenumber associated with the chosen phase-speed threshold for each case.}
\label{fig:AppA_6}
\end{figure}

\begin{figure}[t]
\centering
\includegraphics[width=0.99\textwidth]{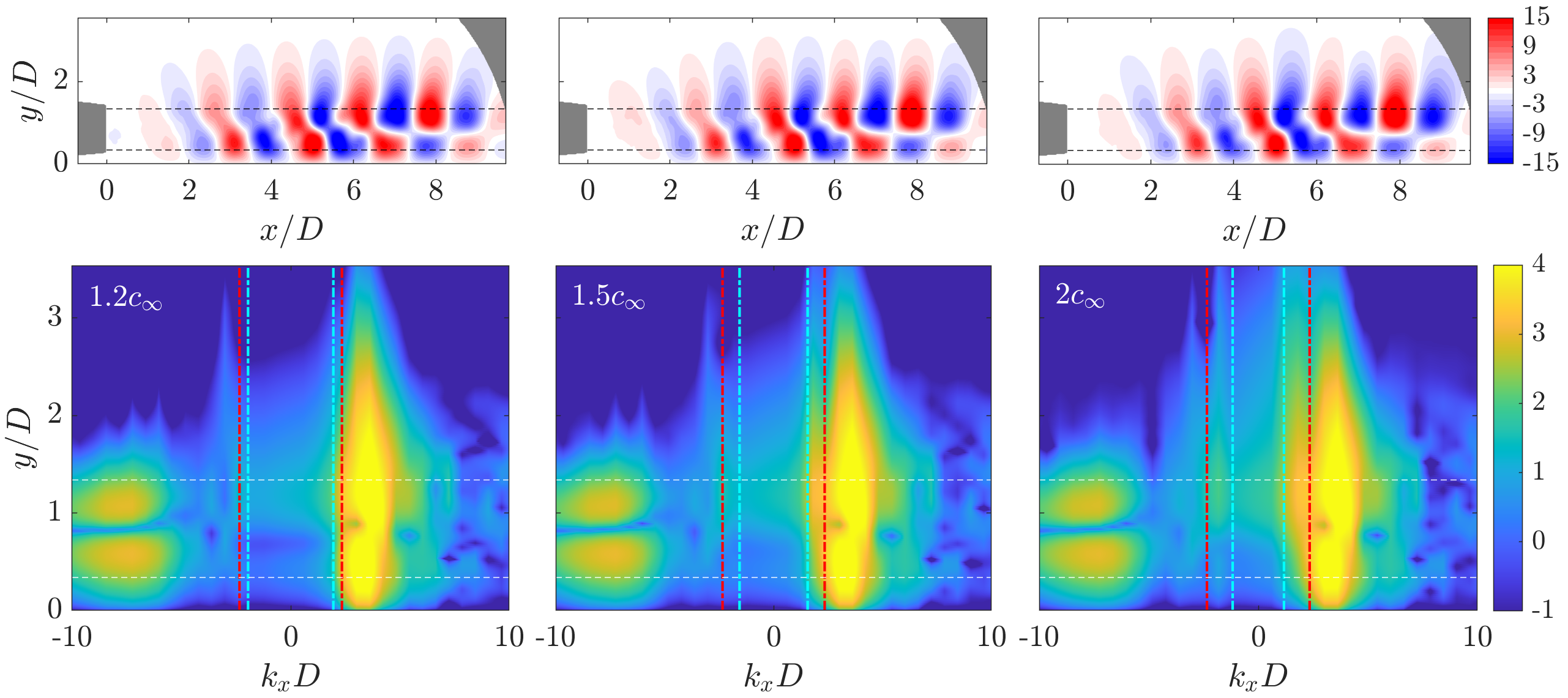}
\caption{Comparison of the second antisymmetric $\Theta'$ SPOD mode for $M_j = 1.26$, $St = 0.36$ using different phase-speed thresholds for removing unphysical harmonic waves: (left) $c_{ph,c} = 1.2 c_\infty$; (middle) $c_{ph,c} = 1.5 c_\infty$; (right) $c_{ph,c} = 2 c_\infty$; (top) real part of the SPOD mode amplitude function; (bottom) logarithm of the energy magnitude from the streamwise Fourier transform of the SPOD mode.}
\label{fig:AppA_7}
\end{figure}

The choice of the phase-speed threshold employed to remove the most energetic harmonic waves can have an impact on the SPOD eigenvalue problem. To assess the sensitivity of the SPOD results with respect to $c_{ph,c}$, three different threshold values are considered, namely $c_{ph,c} = 1.2 c_\infty, 1.5 c_\infty$ and $2 c_\infty$.

Figure \ref{fig:AppA_5} presents a comparison of the antisymmetric SPOD spectra ($M_j = 1.26$) for the first and second SPOD modes obtained using the different filtering thresholds. The unfiltered SPOD spectra are also included for comparison. Two different important observations can be made. On the one hand, the comparison between filtered and unfiltered results shows that the unphysical energy components have a stronger influence at higher frequencies. On the other hand, the filtered results are not very sensitive to the selected phase-speed threshold. Their sensitivity is higher for lower frequencies because the acoustic wavenumber decreases with frequency. For very low frequencies, the unphysical harmonic energy band and the physical energy bands become very close together. Therefore, small values of $c_{ph,c}$ can impact the physical energy bands as well, which explains the small effect shown in the spectra. In particular, at low frequencies, small phase-speed thresholds can have a non-negligible effect on the growth of the coherent structures (wavepackets), which is governed by small wavenumbers dictated by the wavepacket envelope shape. For high frequencies, however, the range of wavenumbers contaminated by the artificial harmonic components is far form the acoustic wavenumber where the physical energy is generally concentrated. As a result, the spectra are very insensitive to the filter width in this regime.

For the results presented in this work, the influence of the employed phase-speed threshold ($1.2 c_\infty$) is negligible. Figures \ref{fig:AppA_6} and \ref{fig:AppA_7} illustrate this by respectively comparing the first and second SPOD modes at $St = 0.36$ (the lowest frequency analyzed in the results section) for the three different filtering thresholds under consideration. The impact of the filter width on the amplitude function and the corresponding energy signature is very small; only the growth rate of the coherent structures of the second SPOD mode shows small deviations with different filter width. For the analysis of lower-frequency coherent structures, however, larger values of $c_{ph,c}$ are recommended.

\section{Comparison of alternative approaches to construct the $\Theta'$ SPOD eigenvalue problem} \label{sec:AppB}

Different approaches can be considered to construct the SPOD eigenvalue problem based on the cross-spectral density of $\Theta'$. Besides the methodology outlined in section \ref{sec:Poisson_solution}, where the Poisson equation is solved in the frequency-wavenumber domain, two other procedures have also been tested, which are summarized below.

\begin{figure}[!ht]
\centering
\includegraphics[width=0.99\textwidth]{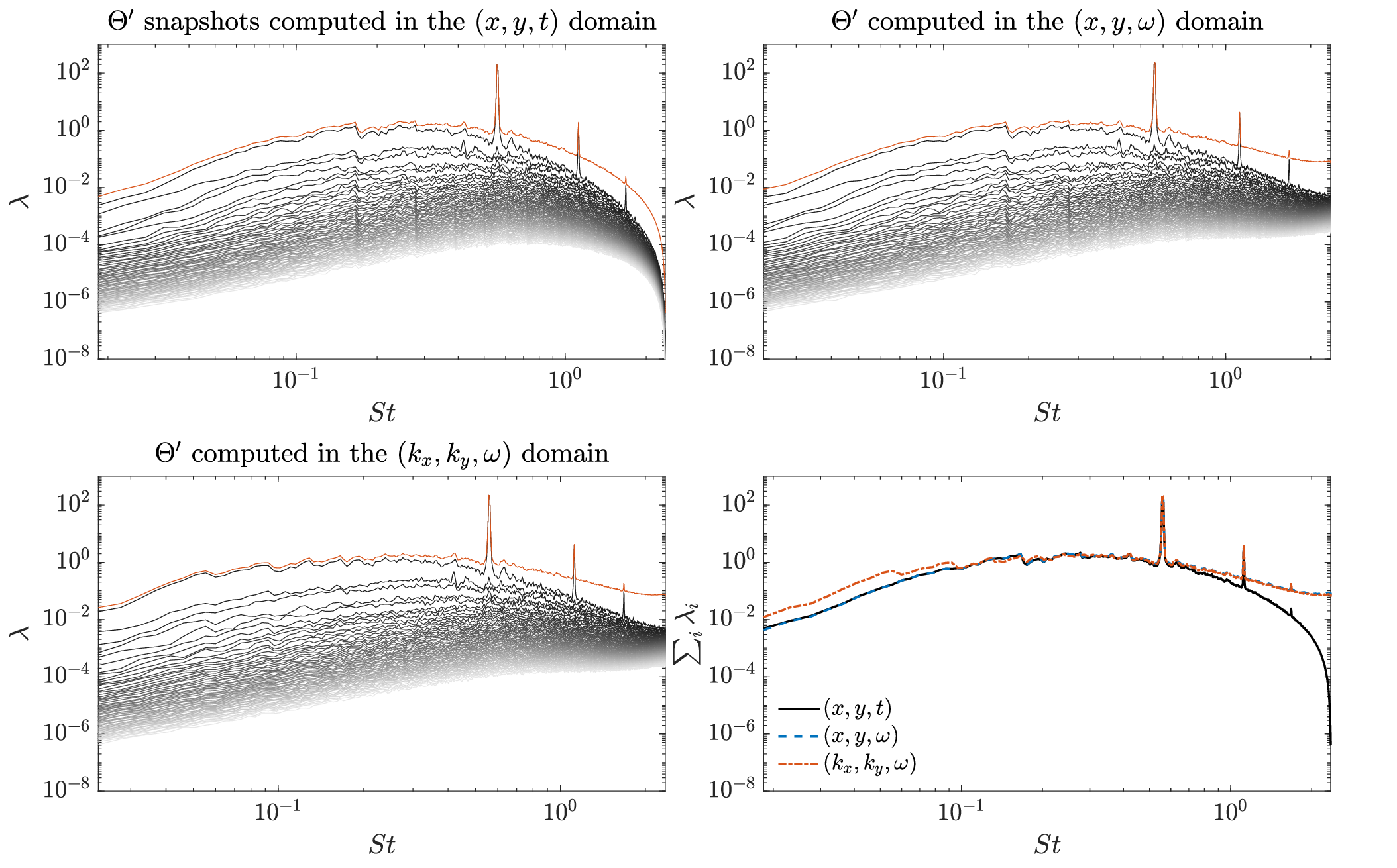}
\caption{Comparison of the symmetric $\Theta'$ SPOD spectra for $M_j = 1.26$ obtained with different approaches: (top-left) using $\Theta'$ snapshots computed in the space-time domain; (top-right) using $\Theta'$ computed in the space-frequency domain; (bottom-left) using $\Theta'$ computed in the wavenumber-frequency domain; (bottom-right) comparison of the sum of the energy of all the SPOD modes for the three approaches.}
\label{fig:AppB_1}
\end{figure}

\subsection{Calculation of $\Theta'$ snapshots in the space-time domain}

In this approach, $\Theta'$ snapshots are directly computed in the space-time domain through the numerical solution of equation \eqref{eq:Poisson_sch}. The time derivative of Schlieren fluctuations (source term of the Poisson equation) is in this case computed by means of classical second-order finite differences. Similarly, the Laplacian operator is discretized using a second-order central finite-difference scheme for a uniform grid. Regarding the boundary conditions, an homogeneous Dirichlet condition is imposed. This results in a linear system of equations which is solved by means of sparse Cholesky factorization using the CHOLMOD package~\cite{Chen2008}. The SPOD algorithm is then applied directly on $\Theta'$ snapshots. Once the realizations of $\Theta'$ are Fourier-transformed in time, the FFT of $\tilde{\Theta}'$ along $x$ is calculated and the most energetic harmonic components are removed in the same way as introduced before.

The symmetric SPOD spectrum resulting from this approach is represented in the top left plot of figure \ref{fig:AppB_1} for $M_j = 1.26$. Although the broadband mixing noise region and the screech resonance tones are well predicted, this method leads to a very fast decay of the SPOD energy for high frequencies. The reason for this lies on the second-order scheme employed for the calculation of the time derivative of the Schlieren fluctuations, which attenuates the high-frequency content present in the original signal. In order to preserve the energy at high frequencies, the time derivative should be computed in the frequency domain or by using a high-order discretization scheme in the time domain. The use of high-order differentiation schemes, however, carries associated computational difficulties such as numerical instabilities.

The application of homogeneous Dirichlet boundary conditions in this approach may be combined with the use of a sponge layer to force the Schlieren fluctuations to decay to zero towards the boundaries and avoid discontinuities (as done in \cite{Prasad2022}). Here, the use of a Planck-taper spatial window has been tested for that purpose. In all the cases investigated, using tapering parameter values of $\varepsilon = 0.1, 0.15$ and $0.2$, no significant reduction of the unphysical harmonic energetic components introduced in the solution could be achieved. Therefore, the filtering process is still required to obtain meaningful coherent structures.

\subsection{Calculation of $\tilde{\Theta}'$ in the space-frequency domain}

In this case, snapshots of $\Theta'$ are not obtained in the time domain. Instead, the calculation of $\tilde{\Theta}'$ is carried out by solving the Poisson equation in the space-frequency domain. This is achieved by computing the time derivative of Schlieren fluctuations in the frequency domain, but solving the resulting Poisson equation \eqref{eq:Poisson_sf} in the $x,y$ spatial domain using a second-order finite-difference scheme, as in the previous approach.

The spectrum obtained via this method is displayed in the top right plot of figure \ref{fig:AppB_1}. With this approach, the obtained spectrum is almost identical to the one obtained solving for $\Theta'$ in the wavenumber-frequency domain. However, it is important to note that the computational cost of this alternative procedure, as well as of the other one presented in this appendix, is significantly higher than the methodology outlined in \S \ref{sec:Poisson_solution}, since they require the solution of a linear system of equations for each $\Theta'$ field and its posterior spatial Fourier transform to remove the unphysical harmonic waves. Therefore, these results justify the choice of the solution strategy adopted in this study.

\end{appendices}

\bibliographystyle{spmpsci}
\bibliography{referencesTCFD}

\end{document}